\documentclass[default]{sn-jnl}% Default
\usepackage[utf8]{inputenc}\usepackage[english]{babel}
\usepackage{subfigure}
\usepackage{comment}

\jyear{2021}%

\begin{document}
\newcommand{\svo}{\textit{SVOM}}
\newcommand{\ca}{CAGIRE}
\newcommand{\nir}{near-infrared}
\newcommand{\arcmin}{$^{\prime}$}
\newcommand{\arcsec}{$^{\prime\prime}$}

\title[CAGIRE]{CAGIRE: a wide-field NIR imager
for the COLIBRI 1.3 meter robotic telescope}

%% Suffix	-> \sfx{IV}s

\author*[1]{\fnm{Alix} \sur{Nouvel de la Flèche}}\email{alix.nouvel-de-la-fleche@irap.omp.eu}

\author[1]{\fnm{Jean-Luc} \sur{Atteia}}
\author[1]{\fnm{Jérémie} \sur{Boy}}
\author[1]{\fnm{Alain} \sur{Klotz}}
\author[1]{\fnm{Arthur} \sur{Langlois}}
\author[1]{\fnm{Marie} \sur{Larrieu}}
\author[1]{\fnm{Romain} \sur{Mathon}}
\author[1]{\fnm{Hervé} \sur{Valentin}}

\author[1]{\fnm{Philippe} \sur{Ambert}}
%\author[1]{\fnm{Francis} \sur{Beigbeder}}
%\author[1]{\fnm{Patrick } \sur{Couderc}}

\author[2]{\fnm{Jean-Claude} \sur{Clemens}}
\author[2]{\fnm{Damien} \sur{Dornic}}
\author[2]{\fnm{Eric} \sur{Kajfasz}}
\author[2]{\fnm{Jean} \sur{Le Graët}}
\author[2]{\fnm{Olivier} \sur{Llido}}
\author[2]{\fnm{Aurélia} \sur{Secroun}}

\author[3]{\fnm{Olivier} \sur{Boulade}}
\author[3]{\fnm{Ayoub} \sur{Bounab}}

\author[4]{\fnm{Giacomo } \sur{Badano}}
\author[4]{\fnm{Olivier} \sur{Gravrand}}

\author[5]{\fnm{Sébastien } \sur{Aufranc}}
\author[5]{\fnm{Adrien} \sur{Lamoure}}
\author[5]{\fnm{Lilian} \sur{Martineau}}
\author[5]{\fnm{Laurent} \sur{Rubaldo}}

\author[6]{\fnm{Hervé} \sur{Geoffray}}
\author[6]{\fnm{François} \sur{Gonzalez}}

\author[7]{\fnm{Stéphane} \sur{Basa}}
\author[7]{\fnm{François } \sur{Dolon}}
\author[7]{\fnm{Johan} \sur{Floriot}}
\author[7]{\fnm{ Simona} \sur{Lombardo}}

\author[8]{\fnm{Salvador} \sur{Cuevas}}
\author[8]{\fnm{Alejandro} \sur{Farah}}
\author[8]{\fnm{Jorge} \sur{Fuentes}}
\author[8]{\fnm{Rosalía} \sur{Langarica}}
\author[8]{\fnm{Alan} \sur{M. Watson}}

\author[9]{\fnm{Nathaniel} \sur{Butler}}

\affil*[1]{\orgname{IRAP,Université de Toulouse, CNRS, CNES, UPS}, \orgaddress{\postcode{31401}, \city{Toulouse}, \country{France}}}

\affil[2]{\orgname{Aix Marseille Univ,CNRS/IN2P3, CPPM}, \orgaddress{\city{Marseille}, \country{France}}}

\affil[3]{\orgname{CEA-IRFU}, \orgaddress{\street{Orme des Merisiers},\postcode{91190} \city{Gif-sur-Yvette}, \country{France}}}

\affil[4]{\orgname{CEA-LETI}, \orgaddress{\street{17 Avenue des Martyrs}, \postcode{38054},\city{Grenoble}, \country{France}}}

\affil[5]{\orgname{LYNRED}, \orgaddress{\street{364 Av. de Valence}, \postcode{38113},\city{Veurey-Voroize}, \country{France}}}

\affil[6]{\orgname{CNES}, \orgaddress{\street{18 Av. Edouard Belin}, \postcode{31401},\city{Toulouse}, \country{France}}}

\affil[7]{\orgname{Aix Marseille Univ, CNRS, CNES, LAM,}, \orgaddress{\city{Marseille}, \country{France}}}

\affil[8]{\orgname{Instituto de Astronomía, Universidad Nacional Autónoma de México} \orgaddress{\street{}, \postcode{Apartado Postal 70-264}, \city{04510 CDMX}, \country{México}}}

\affil[9]{\orgname{School of Earth and Space Exploration, Arizona State University} \orgaddress{\street{}, \postcode{PO Box 871404}, \city{Tempe AZ 85287}, \country{USA}}}


\abstract{The use of high energy transients such as Gamma Ray Bursts (GRBs) as probes of the distant universe relies on the close collaboration between space and ground facilities.
In this context, the Sino-French mission \svo\ has been designed to combine a space and a ground segment and to make the most of their synergy. 
On the ground, the 1.3 meter robotic telescope COLIBRI, jointly developed by France and Mexico, will quickly point the sources detected by the space hard X-ray imager ECLAIRs, in order to detect and localise their visible/NIR counterpart and alert large telescopes in minutes.
COLIBRI is equipped with two visible cameras, called DDRAGO-blue and DDRAGO-red, and an infrared camera, called CAGIRE, designed for the study of high redshift GRBs candidates. 
Being a low-noise NIR camera mounted at the focus of an alt-azimutal robotic telescope imposes specific requirements on CAGIRE.  
We describe here the main characteristics of the camera: its optical, mechanical and electronics architecture, the ALFA detector, and the operation of the camera on the telescope. 
The instrument description is completed by three sections presenting the calibration strategy, an image simulator incorporating known detector effects, and the automatic reduction software for the ramps acquired by the detector.
This paper aims at providing an overview of the instrument before its installation on the telescope. 
}

% Freely adapted from https://www.aanda.org/author-information/information-files/170-aaa-keywords


\maketitle
%%%%%%%%%%%%%%%%%%%%%%%%%%%%%%%%%%%%%%%%%%%%%%%%%
%début texte 
%%%%%%%%%%%%%%%%%%%%%%%%%%%%%%%%%%%%%%%%%%%%%%%%%

\section{Introduction }\label{sec:intro}

Using Gamma Ray Bursts (GRBs) as probes of the distant universe involves the quick and precise localisation of the bursts, the multi-wavelength observation and high resolution spectroscopy of their afterglows and the detailed imaging and spectroscopy of their host galaxies. Such observing sequences require the efficient cooperation of space and ground facilities. In this context the \svo\ mission (Space based multi-band astronomy Variable Object Monitor, \cite{Wei2014, Atteia2022}) includes two follow-up telescopes designed to quickly localise and monitor the optical afterglows of GRBs detected by the space instruments. Among them, the COLIBRI telescope  (Catching Optical Light and Infrared BRIght transients, \cite{Basa2022}), equipped with the near-infrared CAGIRE camera (CApturing Gamma-ray bursts Infra-Red Emission), plays a special role for the detection and localisation of the most distant GRBs, at redshift z $\ge 6.5$, whose emission is outside the range of the Visible Telescope (VT) of \svo\ \citep{Wu2012,Fan2020}.

Despite the scarcity of such very distant GRBs, their observation encompasses key objectives that have been considered sufficiently important to justify the development of a dedicated near-infrared camera at the focus of a mid-size robotic telescope.
Improving our capabilities to observe high-z GRBs CAGIRE will hopefully contribute to constrain the GRB rate beyond redshift 5-6 \citep{Daigne2006, Kistler2009}, facilitate the detection of the faint hosts of very distant GRBs \citep{Basa2012, Tanvir2016} and eventually detect GRBs associated with the explosion of population III stars \citep{Bromm2006, Toma2011}. 

%For GRBs up to redshift z $\approx$ 11, \ca\ on COLIBRI provides a crucial link between the localizations of ECLAIRs, the hard X-ray imager and trigger of \svo\ with a location accuracy of few arcminutes \citep{Godet2014}, and sub-arcsecond localizations needed to perform high resolution spectroscopy of GRB afterglows with large ground-based telescopes.

\subsection{CAGIRE in the context of \svo }
\label{sub:svo}
As soon as a GRB will be localised by ECLAIRs, its position will be sent to the platform to slew the \svo satellite, and simultaneously broadcasted to the ground by an on-board  Very High Frequency (VHF) emitter and via a Beidou short message. On Earth, the alert should be detected by one of the $\sim 50$ \svo VHF receiving stations located in the tropical zones. Once the message is received, the VHF station convey it to the French Scientific Center, which redistributes it to the world. 
65\% of the alerts are received on Earth within 30s, allowing ground-based observations to start most of the time before the end of the satellite slew \citep{Wei2014}.

One specificity of \svo\ is the association of a satellite with a strong ground segment that includes a set of  Ground-based Wide Angle Cameras (GWAC, \cite{Han2021}) monitoring the visible sky simultaneously with the satellite and two medium size Ground Follow-up Telescopes:  the Chinese-Ground Follow-up Telescope (C-GFT), and the French-Ground Follow-up Telescope (F-GFT), aka COLIBRI \citep{Basa2022}, respectively located in China and Mexico.

We focus here on the Franco-Mexican robotic telescope COLIBRI, operating from the Observatorio Astronómico Nacional in San Pedro Mártir. The telescope is designed to quickly point transient sources upon alert and to make images of the error box at near infrared and visible wavelengths. The specificity of COLIBRI is the availability of an infrared channel equipped with a camera, called CAGIRE, housing a low noise large area European detector, called ALFA. 

\subsection{CAGIRE performances overview}

CAGIRE is a 2k $\times$ 2k scientific camera sensitive in the near infrared, covering wavelengths from 1.1\,µm to 1.8\,µm. At the focus of the COLIBRI telescope, the camera will acquire images in two photometric channels (J \& H) within a square field of view of 21.7 arcmin on a side, to cover the error boxes of ECLAIRs.  
Before describing the camera, we provide a summary of its main characteristics and performance in table \ref{tab:CAGIRE_Perf}.

The main goal of this paper is to provide useful reference material for the potential users of CAGIRE data. It is organised as follows: section \ref{sec:archi} describes the architecture of the camera, its optics, mechanics and electronics equipment, section \ref{sec:alfa} presents the detector, and section \ref{sec:operations} the operations of the instrument. 
The calibrations and characterisations are briefly discussed in section \ref{sec:carac}, and  
section \ref{sec:simu} presents an image simulator incorporating known detector effects.
Finally, sections \ref{sec:preproc} and \ref{sec:pvalid} respectively introduce the automatic data preprocessing and its validation with simulated ramps. 

\begin{table}[h]
    \begin{minipage}{\textwidth} \centering
    \begin{tabular}{|l l|}
         \hline
         \textbf{Parameter }& \textbf{Value}  \\
         \hline
         \multicolumn{2}{|c|}{\textbf{CAGIRE parameters}} \\
         \hline
         Field of view & 21.7'\\
         \hline
         Sky pixel size & 0.65''\\
         \hline
         Wavelength range  & 1.1µm to 1.8µm\\
         Photometric channels & J and H\\
         \hline
         Sky background expected & 160 e-/s/pix (J)\\
                                & 1250 e-/s/pix (H)\\
         \hline
         Maximum attainable redshift size & z $\sim$ 11\\
         \hline
         Vacuum autonomy & $>$6 months\\
         \hline
         Image processing duration  (time since trigger) & $<$5 min\\
         \hline
         \multicolumn{2}{|c|}{\textbf{ALFA detector parameters (see also section \ref{sub:carac})}} \\
         \hline
         Operating temperature & 100K \\
         \hline
         Readout noise $^1$ & 40 e- \\
         \hline
         Dark current & 0.004 e-/s/pix at 100K \\
         \hline
         Diode dynamic range $^2$& 240 ke- \\
        \hline
        Quantum efficiency & 60 \% \\
        \hline
        Inter Pixel Capacitance (IPC) & 0.8\% \\
        \hline
        Conversion gain  & $\sim 10$ e-/ADU \\
        \hline
        Operability under flux & 99 \% \\
        \hline
    \end{tabular}
    \end{minipage}
    $^1$ {The readout noise presented here is computed from CDS noise, and has been measured after correction by the reference pixels, see \citep{NouvelDeLaFleche2022}. Without correction, we find a readout noise of 55e-.}\\
    $^2$ {The dynamic presented here is the full dynamic of the diode, measured with CAGIRE configuration polarisation of the diode, $\Delta V_{diode}$=0.6V. }
    \caption{Summary of CAGIRE performance}
    \label{tab:CAGIRE_Perf}
\end{table}

\section{Instrument architecture}
\label{sec:archi}

CAGIRE is composed of three main subsystems (fig. \ref{fig:CagireOnColibri}): a cryostat behind a filter selector (CAGIRE in fig. \ref{fig:CagireOnColibri}), a close electronics, both onboard the telescope, and a remote electronics located in the control room. We provide below more details on the instrument's architecture. 

\subsection{Design tradeoffs}
\label{sub:design}
Using a cooled NIR camera at the Nasmyth focus of an Alt-Azimutal telescope, like COLIBRI, requires specific tradeoffs discussed in this section.

The most important requirement on CAGIRE is to cover ECLAIRs error boxes in a single exposure with a 2k $\times$ 2k detector. The 21.7\arcmin\ field of view of CAGIRE sets the size of the sky pixel to 0.65\arcsec\ on a side. This determines the expected sky signal, which is expected to reach $160$ and $1250~\mathrm{e^-/s/pix}$ in channels J and H respectively. 
As a consequence of the optical design described in section \ref{sub:optic}, the thermal radiation contributes to only a small fraction of the sky signal and there is no need to use cold optics for CAGIRE. 
The instrument has been designed with the aim of being sky-limited, in the sense that the main source of background are the Poisson fluctuations of the sky signal. The instrument internal background, for instance, is below 25\% of the sky signal, and the detector readout noise is smaller than the fluctuations of the sky signal for exposures longer than 20 seconds in channel J and 2 seconds in channel H. These values are shown in table \ref{tab:expected_signal}

\begin{table}[h]
    \centering
    \begin{tabular}{|c|c|c|c||c|c|c|}
    \hline
    channel and time    & J (1s) & J (20s) & J (60s) & H (1s) & H (20s) & H (60s) \\
    \hline
    sky  &  160     &   3 200     &   9 600     &  1 250    &   25 000     &  75 000    \\
    \hline
    internal signal &   8   &   160    &   480    &  17.3    &   346     &    1 038   \\
    \hline
    \hline
    sky fluctuation  & \textbf{13}       &  57      & 98     &  \textbf{35}     &  158     &    273    \\
    \hline
    internal signal variation &  3  &  13     &   23    &  4   &   18     &   32    \\
    \hline
    readout noise &     \multicolumn{6}{|c|}{40 }        \\
    \hline
    \hline
    Total noise &  42  &   71    &  108     &   54   &   164     &   279    \\
    \hline
    \end{tabular}
    \caption{Expected signal and fluctuations en e-/pix. Numbers in bold indicate cases where the readout noise exceeds the sky fluctuations}
    \label{tab:expected_signal}
\end{table}

A second requirement concerns the autonomy of the instrument: CAGIRE should work with less than one maintenance period per 6 months (with a goal of 1 per year). 
To achieve this goal the instrument uses a cryocooler that is continuously running to keep the operating temperature below 100~K and relies on passive cryo-pumping to maintain the vacuum.
With an aim of reliability, the instrument has few moving parts: two translation stages, a warm shutter and a linear motion stage. The warm shutter is closed during the telescope repointing, in order to prevent bright stars from illuminating the detector. The first translation stage is used to adjust the focus by moving a lens of the warm optical bench. The second one moves the filter slide in front of the cryostat. The linear motion stage is used to push a cold shutter in front of the detector, isolating it from the instrument and sky signals, allowing to monitor its health in a controlled environment.

\begin{figure}[h]
    \centering
    \includegraphics[width=8cm]{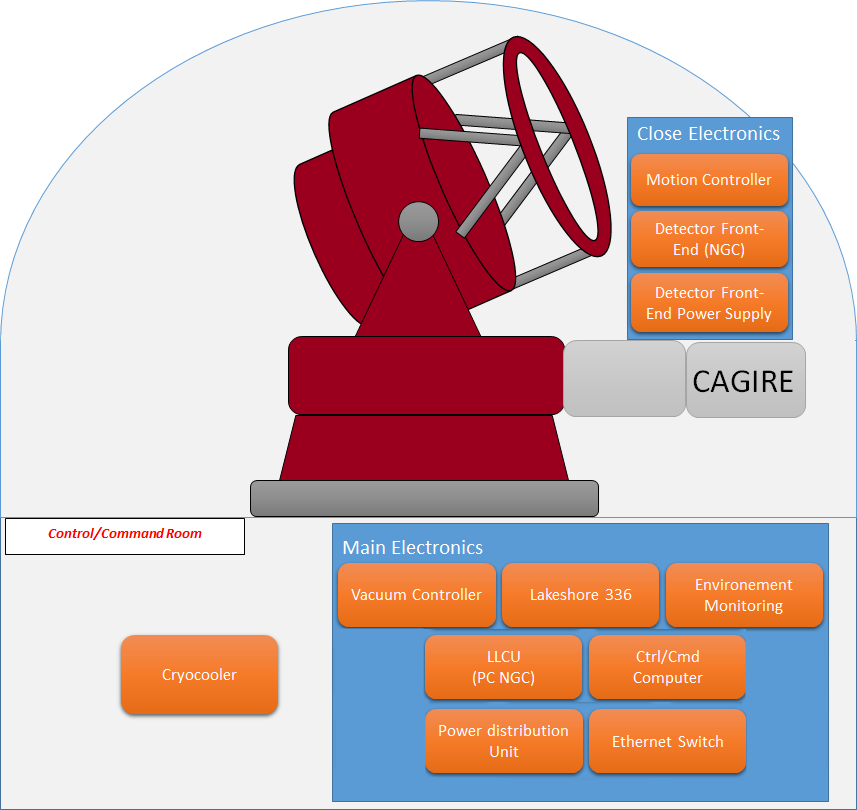}
    \caption{Schematic diagram of the CAGIRE camera}
    \label{fig:CagireOnColibri}
\end{figure}

Being at the focus of an Alt-Az telescope, the camera will always move (during and between the observations), driven by the mechanical derotator of the telescope. As the derotator cannot support the full weight of the camera with its electronics, the instrument is divided into two parts: the ``onboard elements'' and the distant elements, as shown in figure \ref{fig:CagireOnColibri}. 
The onboard elements are mounted on the main structural unit (hereafter MSU) attached to the derotator. They include the cryostat, and the close electronics, encompassing the detector front-end electronics and the motion controller. The distant elements lie in a dedicated room and are connected to the onboard elements through a cable wrap.

Another tradeoff involved the choice of the wavelength range of CAGIRE. In order to make the instrument as simple as possible, the wavelength range has been limited to the  J \& H photometric channels.
This choice implies that the y channel is covered by the red CCD \citep{Watson2018}, which has a quantum efficiency (QE) slightly lower than the QE of the NIR detector.

Finally, CAGIRE uses a detector loaned by ESA (see Section \ref{sec:alfa}), which imposed other constraints on the instrument. First, the detector and its cryogenic preamplifier developed by the Commissariat à l'Energie Atomique et aux énergies alternatives (CEA) and inspired from an ESO preamplifier, must be housed inside the cryostat. Second, the detector control and readout is done with ESO's ``New General detector Controller'' (hereafter NGC \cite{Stegmeier2008, Baade2009}), which must not be farther than 3 meters from the detector and must be onboard the telescope.

The requirement to identify afterglow candidates in less than 5 minutes puts no specific constraint on the instrument itself, but it has an impact on its operation as discussed in Section \ref{sec:operations}.

\subsection{Optical design} 
\label{sub:optic}

\begin{figure}[h]
    \centering
    \includegraphics[width=8cm]{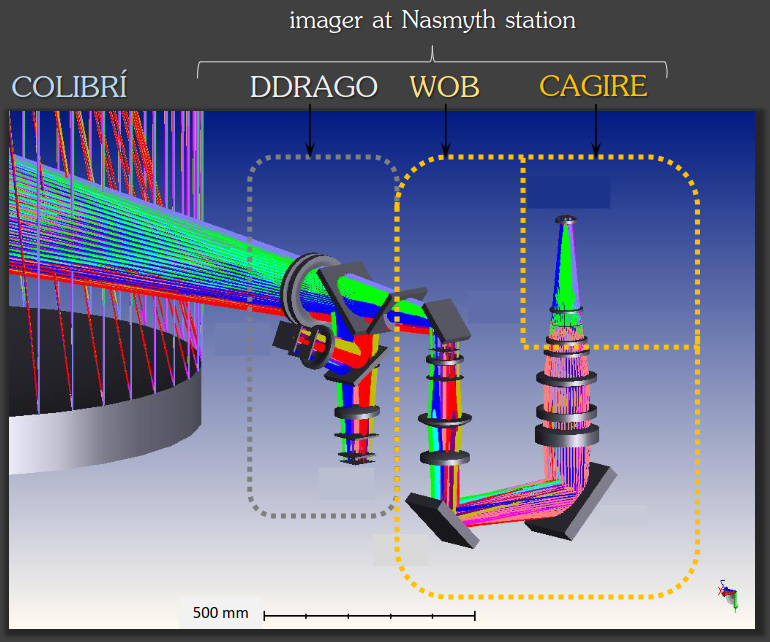}
    \caption{Optical path to CAGIRE, internal document, Jorge Fuentes and Rosalia Langarica}
    \label{fig:optical_path}
\end{figure}

\begin{figure}[h]
    \centering
    \includegraphics[width=12cm]{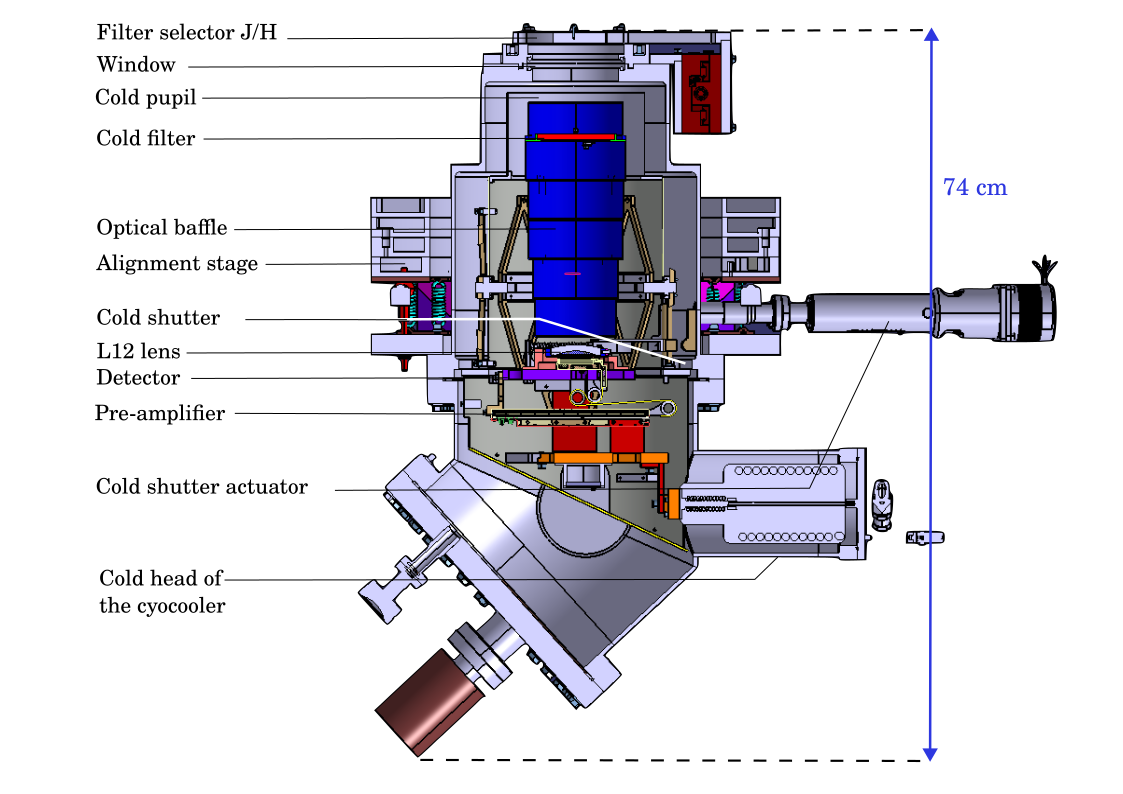}
    \caption{Layout of the CAGIRE camera}
    \label{fig:cagire}
\end{figure}

As illustrated in Fig. \ref{fig:optical_path}, the light arriving from the telescope goes through the Warm Optical Bench (WOB), described in \cite{Fuentes2020}, \cite{Farah2022} and \cite{Langarica2022} before entering the cryostat. The WOB focuses the light onto the CAGIRE detector and eventually blocks the light beam when the telescope is moving. The light then crosses a warm silica filter selecting the desired photometric channel: J or H. A filter selector attached to the cryostat allows to switch between the two filters in less than 5~seconds. 
Inside the cryostat, a cold pupil stops the light coming from the warm objects located outside of the telescope beam, and a fixed cold filter cuts all wavelengths longer than 1.8\,µm.
The light finally crosses a re-imaging cold lens (L12 in figure \ref{fig:cagire}) before getting to the detector. Carefully designed optical baffles prevent stray light to reach the detector.
If needed, a cold shutter can be can be placed in front of the detector to put it into darkness. It is used during engineering periods and not during normal operations. All these elements are presented figure \ref{fig:cagire}. 

\subsection{Mechanical structure}

The WOB, the cryostat, and the close electronics are mounted on the MSU \citep{Langarica2022}. 
The cryostat is mounted on the MSU thanks to a removable plate and an alignment stage allowing to adjust the camera position and orientation along 5 degree of freedom (see figure \ref{fig:cagire}). 
The close electronics includes the detector front-end and its power supply, the motion controllers for the filter selector and the cold shutter, and various environmental probes, placed inside a standard 600~mm wide electronic rack.
The cryocooler and the remote electronics are located in the control room, and described in section \ref{ssub:ControlElec}.

The cryostat with all its parts will be assembled beforehand and then mounted on the alignment stage. Various analyses of the alignment stage and the cryostat have been conducted. They include a static analysis to estimate the impact of gravity on the detector's position inside the cryostat, and a vibration modes analysis in order to minimise the response of the camera to the cryocooler vibrations.
The displacements measured during these tests are smaller than the optical tolerance, with a maximum lateral displacement of $\pm 225~ \mathrm{\mu m}$, smaller than the requirement of $\pm 360~ \mathrm{\mu m}$. 

Additional studies involved the displacement of the detector during the movements of the telescope and a thermo-mechanical analysis to simulate the stress on the structure and the changes of the position of key elements due to thermal expansion/contraction. A maximal contraction of $\pm 300~ \mathrm{\mu m}$ has been observed for some non-optical parts like the pre-amplifier, because of their material, but these parts are mounted on flexible attachments which will absorb the deformation. No significant stress is to be expected in the assembly due to temperature.

\subsection{Cryostat}
\label{sub:cryostat}

\begin{figure}[h]
    \centering
    \includegraphics[width=11cm]{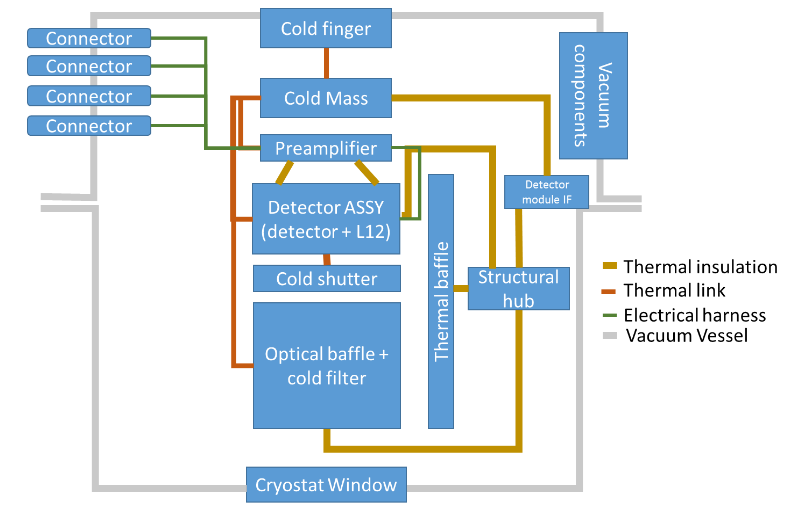}
    \caption{Thermal and electrical model of the CAGIRE cryostat}
    \label{fig:cryostat}
\end{figure}

As the detector selected for CAGIRE operates at a temperature of 100 K, a cryostat is needed to ensure nominal performances. However, the location of the camera on a fast-moving alt-az robotic telescope leads to some constraints, such as the need to provide passive cryogenic vacuum for at least 6 months without pumping, or the rigidity and mass budget of the cryostat. The cryostat, schematically presented in fig. \ref{fig:cryostat}, is composed of a vacuum vessel in aluminium, which houses the optical baffles, the cold filter, the cold shutter, the L12 lens, the detector and the pre-amplifier. It is fitted with a window to let the light enter. 
All the cold components are thermally linked to the cold finger connected to the cold head, itself connected to the cryocooler located in the control room thanks to 38 meter long gas pipes. This allows to cool the cryostat to below 100K, with low vibrations. 
Various connectors, positioned on a single plate, allow electrical connections to the warm electronics. 
The cryostat also hosts temperature and vacuum probes, as explained in section  \ref{ssub:ControlElec}. 

\subsection{Electronics}
\label{sub:elec}

\subsubsection{Acquisition chain}
\label{ssub:acqChaine}

\begin{figure}[h]
    \centering
    \includegraphics[width=\columnwidth]{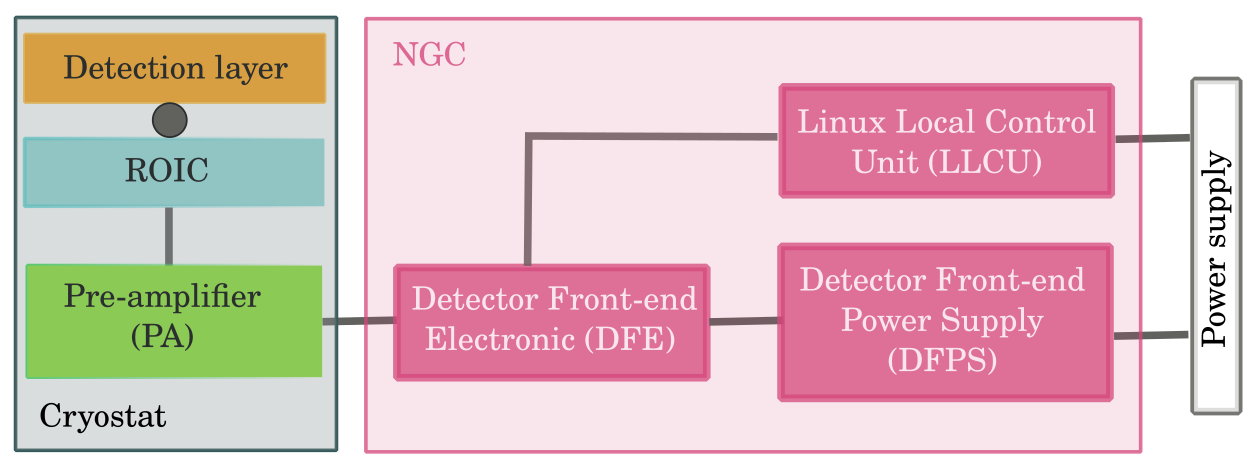}
    \caption{Detection chain diagram. As presented in figure \ref{fig:CagireOnColibri}, DFE and DFPS are mounted on the telescope while the LLCU is located in the control room, below the telescope}
    \label{fig:NGC}
\end{figure}

The detector is controlled by the New General Controler (NGC), developed by the European Space Organisation (ESO) \citep{Stegmeier2008}. The NGC is composed of a Detector Front-end Electronics (DFE) powered by a Detector Front-end Power Supply (DFPS), both located in the close electronic rack, and of a control computer, the Linux Local Control Unit (LLCU), located in the control room and linked to the DFE with an optical fiber. The DFE encompasses various electronic boards, with specific roles to provide the clocks and voltages required to configure and read the detector (CLDC: Clock and DC voltage generator) and to convert the analog video signals from the 32 output channels of the pre-amplifier into numerical values encoded over 16 bits (ADC: Analogue to Digital Converter). The clock patterns and biases levels can be defined by the user in dedicated configuration files. 
Acquisition requests go through the Linux Local Control Unit (LLCU) and are translated for the detector by the Detector Control Software (DCS). What is called hereafter the detection chain is thus composed of the detector (detection layer and readout circuit), the pre-amplifier (PA) and the NGC. 
A diagram of the detection chain is given fig.~\ref{fig:NGC}. 
The preamplifier (PA) has been developed by the Institut de Recherche sur les lois Fondamentales de l'Univers (CEA-IRFU). It has a gain of 4 and in normal science operation this adds no significant noise to the signal. The PA can operate from $\sim$50K to $\sim$ 300K.

\subsubsection{Motion and environment controls}
\label{ssub:ControlElec}

CAGIRE electronics includes various subsystems that control the environment or move the components of the camera. They are listed below, and are represented in the electronics architecture diagram in annex \ref{sub:A1}:

\begin{itemize}
    \item Environmental probes include temperature, humidity, barometric pressure and luminosity sensors from Embedded Data Systems.  These probes, located on the cryostat and in the close electronics rack, are connected, through a "1-Wire server", to the environmental monitoring server located in the the control room.
    \item The detector and the pre-amplifier are maintained at a low temperature with a cold head connected to a cyrocooler located in the control room. The cryostat internal temperature is monitored to better than 0.01 K by a controller from the Lakeshore company.     \item Pressure measurements are done with a Pirani gauge with a cold cathode able to measure pressure over a broad range, from $10^{-9}$ hPa to $10^{3}$ hPa. This gauge is read through a vacuum controller located in the main electronics rack, in the control room. Its data are collected via RS485 communication, and then converted to go through the Ethernet network. 
    \item The filter selector is a unidirectional motorised stage mounted on the cryostat. It has a repeatability of 0.5\,µm, which is needed to reposition the filters at the same position each time. To meet COLIBRI requirements, it is also fast, with a speed of 50\,mm/s, allowing to change the filter in about 2 seconds. It is controlled by a motion controller located in the close electronics rack, on the telescope.
    \item The cold shutter is moved by an actuator located on the cryostat. It is controlled by a motion controller located in the close electronics rack, on the telescope.
    \item The rack in the control room also houses the LLCU (NGC computer, see \ref{ssub:acqChaine}), the inverter generator (UPS) from Schneider company, and the power distribution unit. 
    \item The cryocooler is located next to the rack, still in the control room.
\end{itemize}

\subsection{CAGIRE software}
\label{sub:software}

The CAGIRE software aims to set the camera in the observing condition requested by the Telescope Control System (TCS).
To achieve this objective, the  CAGIRE software is able to: 
\begin{itemize}
    \item Check the environmental conditions outside the cryostat (temperature, humidity, pressure, luminosity).
    \item Cool the cryostat.
    \item Monitor the vacuum.
    \item Ensure the detector security (power and reset configuration).
    \item Control the motorised stages (Filter selector, cold shutter).
    \item Initialise the detector and the detection chain.
    \item Set the acquisition configuration required by the TCS (e.g. number of ramps).
\end{itemize}

\begin{figure}[h]
    \centering
    \includegraphics[width=\columnwidth]{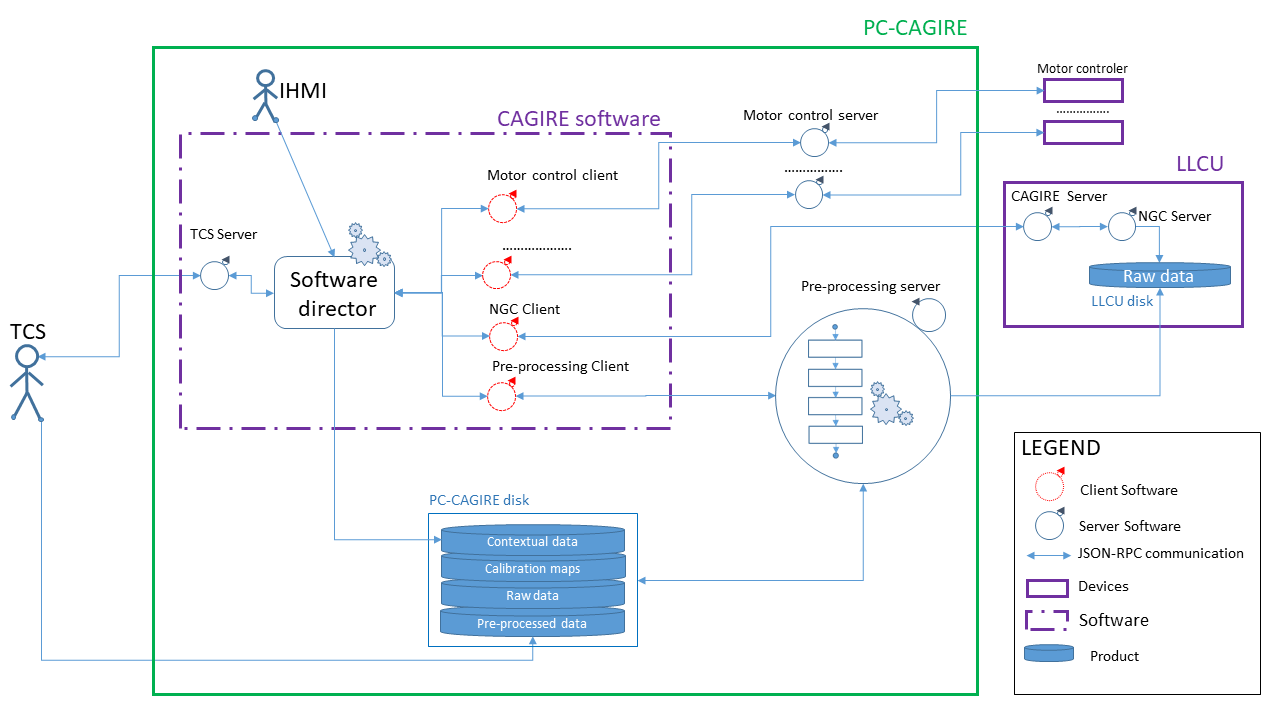}
    \caption{Software architecture. The CAGIRE software is located in the CAGIRE computer and interacts with the severs of various devices }
    \label{fig:software}
\end{figure}

To complete these actions, the software is able to communicate with various devices and to drive them as necessary.
These communications use the JSON-RPC 2.0 protocol, based on JSON files, the software being the client of various servers associated with different devices, as shown in figure \ref{fig:software}. 
When CAGIRE is in its operational configuration, the instrument is ready to respond to requests from the TCS. 
The communication goes through the CAGIRE software, which is the only link between the instrument and the TCS. 
The CAGIRE software, developed in JAVA language, is also equipped with a visible human interface, the IHMI (Interface Homme-Machine Ingénierie or Engineering Man-Machine Interface), giving a control on the different features listed at the beginning of this section. Figure \ref{fig:software} provides a diagram of the CAGIRE software, running in the CAGIRE computer, and its connections to the various devices. 

The devices monitored by the software are either hardware equipment (LLCU, environmental probes, motorised stages) or software developed in JULIA language, like the various device servers and the Pre-processing pipeline. This pipeline will be presented section \ref{sec:preproc}.

Finally, when CAGIRE receives a request from the TCS or through the IHMI, the software checks the status of the sub-systems (motors, environmental probes, LLCU), launches the acquisition and, at the end of the acquisition, launches the preprocessing of the ramp. The raw data and the pre-processed images are then stored and made available to the TCS.

\section{The ALFA detector}
\label{sec:alfa}

CAGIRE uses a new 2k x 2k European detector with low noise and high quantum efficiency in the short infrared wavelength (SWIR), called ALFA, for ``Astronomy Large Format Array'' \citep{Weber2019}. 
The ALFA program, initiated in 2010 and funded by ESA, is supported by the ANR labex FOCUS (Agence Nationale de Recherche Française), and aims at responding to the need of developing a European large sensor array in Near Infra-Red (NIR) for space astronomical applications \citep{Gravrand2022}. 
The first batch of four ALFA sensors included a sensor with excellent performance which has been selected to equip the CAGIRE camera \citep{Gravrand2022}.

\begin{figure}[h]
    \centering
    \includegraphics[width=\columnwidth]{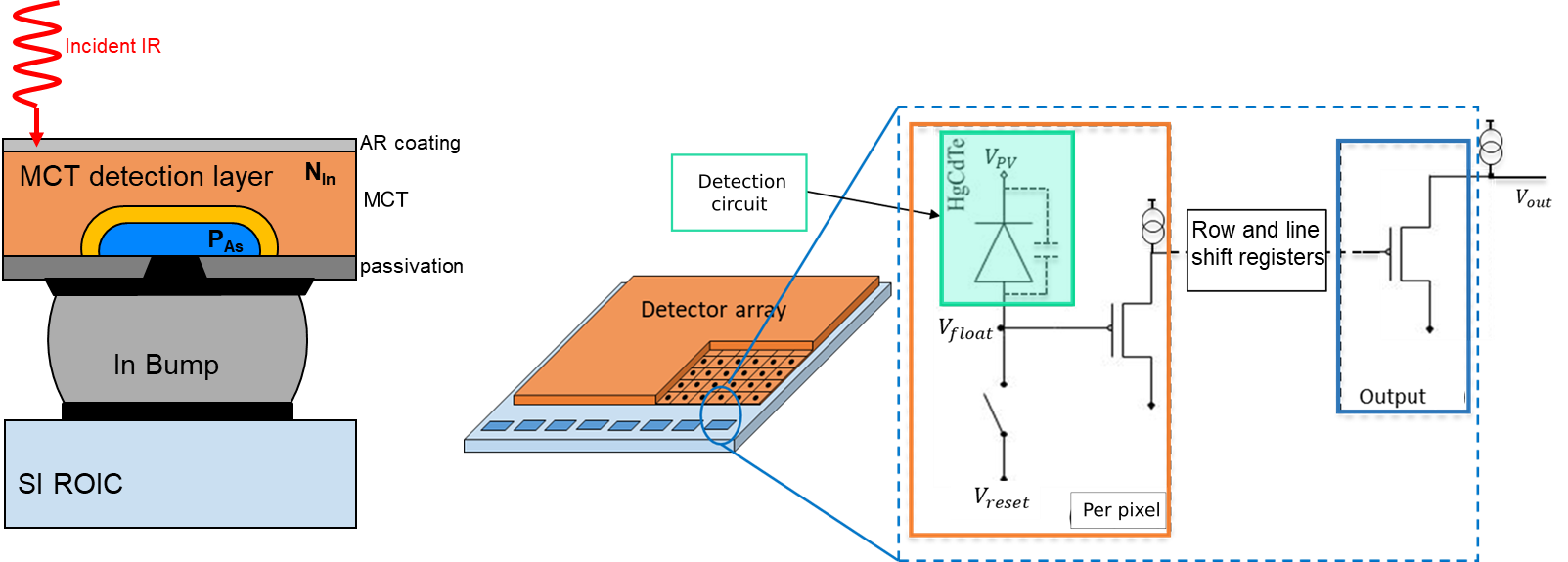}
    \caption{ALFA ROIC and diode schematics - adapted from \cite{Gravrand2022}}
    \label{fig:SFD}
\end{figure}

The Astronomical Large Format Array (ALFA) sensor is an hybrid detector, based on the MCT (Mercury Cadmium Telluride) technology developed at CEA-LETI (fig.\ref{fig:SFD}, left). 
The detection layer is hybridised thanks to indium bumps on a Read-Out Integrated Circuit (ROIC) manufactured by the French company LYNRED. 
The readout is performed through a Source Follower per Detector (SFD) readout circuit, allowing to read very low signals while continuously accumulating charges. 
The sensor is composed of 2048 by 2048 pixels of 15 µm pitch, of which 2040 $\times$ 2040 are active pixels. These active pixels are surrounded by a ring of pixels under permanent reset and a ring of reference pixels, with the same electronics as the active ones, but not sensitive to light. 
The latter can be used for ROIC noise reduction and offset correction. 
Finally, some test pixels with fixed capacitance can be used, for instance, to evaluate the capacitance of active pixels.

The detector covers a spectral range extending from 0.8 to 2.1\,µm \citep{Gravrand2022}. This range covers the J and H photometric bands, but in the case of CAGIRE, it justifies the need of a cold filter to block wavelengths longer than 1.8\,µm that would otherwise increase unwanted signal on the detector. 
The sensor needs to be cooled at 100\,K, and thus will be located into a cryostat. 
The ALFA sensor, because of its SFD structure (see fig.\ref{fig:SFD}), allows to work in a non-destructive mode, called "Up the Ramp". It means that the amount of charges accumulated in the pixels can be read many times between two resets, building ramps of signal whose slope is proportional to the flux received by the pixel. 

Among the various readout modes available, CAGIRE will use the 32-channel science mode, with a 100 kHz readout. In this mode, each pixel is read every 1.33\,s, providing a temporal resolution of 1.33\,s. This mode also enables more flexibility to process the data after their acquisition, and does not lead to data rate issues because of the localisation of the telescope on ground. It nevertheless requires a special management of the data to get astronomical images, this is the role of the preprocessing pipeline described in section \ref{sec:preproc}.

The polarisation voltages selected for the detector combined with the detection gain and the gain of the preamplifier give a conversion factor of about 10 e-/ADU. 
The Analog to Digital Conversion is encoded over 16 bits. 
The diode full dynamic, of $\sim$ 240 000 electrons, is encoded on 24\,kADU, with a typical signal between $\sim$29.5 and $\sim$53.6\,kADU, representing respectively the median bias value and the median saturation level. We notice a linearity better than 5\% over the first 70\% of this range.
More detailed performances of the detector can be found in section \ref{sub:carac} and in \citep{Fieque2018}.

\section{Operations}
\label{sec:operations}

CAGIRE is designed to be an instrument that is both autonomous and simple to operate.
The autonomy relies on two features: first, CAGIRE monitors its own status and is able to switch off automatically and go to safe mode in case of a problem, and second, the operation of the instrument is sufficiently simple to be fully automated.

This choice of simplicity started with the decision to use a single readout mode, the science mode with 32 output channels, which drives the observing mode of CAGIRE.
Upon reception of a request from the Telescope Control System (TCS), CAGIRE executes the following sequence : select the filter, configure the acquisition with the requested number of frames, start and stop the acquisition, preprocess the data and make them available to the TCS, and wait for the next request. 
This sequence can be interrupted at any time by the TCS, for instance upon arrival of a high-priority alert.
Upon request, CAGIRE can also send its current status to the TCS. 

CAGIRE is thus a ``passive instrument'', with no memory and no context (beyond its own configuration status).
No memory means that CAGIRE deals only with the current exposure, keeping no memory of previous exposures (to the exception of the map of saturated pixels used for the management of short-term persistence, as explained in section \ref{sec:preproc}). 
No context means that CAGIRE uses no contextual information, like the pointing direction, the type of exposure (flat or sky), or the dithering strategy, when doing an observation. 
In the end, the complexity is transferred to the TCS, which is in charge of commanding the observations, and to the astronomy pipeline, which is in charge of data analysis. As an example, the dithering strategy will be managed (1) by the TCS to plan the observations and (2) by the astronomy pipeline to reconstruct the images. In this way, a dithering sequence has no impact on CAGIRE observations.

\section{Characterisation }
\label{sec:carac}

Getting the best from a detector like ALFA requires a detailed knowledge of its properties. This is the goal of detector characterisation.
In the case of CAGIRE, the characterisation of the camera is divided into three phases. 

First, the CEA/IRFU performed a precise characterisation of the detector for ESA, who is the proprietary of the detector. 
In a second phase, the CAGIRE detection chain will be tested at the Centre de Physique des Particules de Marseille (CPPM), using a configuration representative of CAGIRE in terms of voltage biases, exposure times or illumination levels. This phase has two mains goals: evaluate the detector response in a controlled environment and in CAGIRE configuration, and acquire the different maps  needed for the preprocessing of images (see section \ref{sec:preproc}). The third and last phase is the characterisation of the fully integrated camera at the Institut de Recherche en Astrophysique et Planétologie (IRAP).

\subsection{Detector characterization at CEA-IRFU} \label{sub:carac}

As the detector is ESA owned, the agency contracted CEA-IRFU/DAp to characterise the sensor according to ESA specifications. 
CEA developed different benches presented in \citep{Pichon2022b}, to measure the characteristics required by ESA. Among them, the dark current, readout noise, linearity, conversion gain, Inter-Pixel capacitance (IPC) and Quantum Efficiency (QE) have been measured on the ALFA detector selected for CAGIRE. 

Table \ref{tab:test_IRFU} shows some important parameters measured at CEA, on the detector selected for CAGIRE. Except for the quantum efficiency, which is slightly lower than expected, these measurements are compliant with CAGIRE requirements. The dark current is negligible compared to other noise sources, IPC presents the expected values, and the linear dynamic range is large enough to avoid saturating the camera too quickly. The readout noise is higher than expected but this is not an issue since it becomes smaller than the Poisson noise of the sky signal for exposures longer than 17\,s in J and 2\,s in H. Furthermore, this value will be reduced by $\sim$ 20-25\% after removing the common mode noise with the reference pixels \citep{Kubik2014}.

The interested reader can find more information on these measurements in \citep{Gravrand2022}, \citep{Pichon2022b} and \citep{Pichon2022}.

\begin{table}[h]
   \begin{minipage}{\textwidth} \centering
    \begin{tabular}{|l l| }
         \hline
         \textbf{Parameter }& \textbf{Mean value over } \\
         \                  & \textbf{ the detector}  \\
         \hline
         Dark current & 0.004 e-/s/pix at 100K \\
         \hline
         Readout noise$^1$ & 40 e- \\
         \hline
         Diode dynamic range $^2$ &  240 ke-\\
        \hline
        Quantum efficiency & 60 \% \\
        \hline
        Inter Pixel Capacitance (IPC) & 2.3 \%  \\
        \hline
        Conversion gain  & $\sim 10$ e-/ADU  \\
        \hline
        Operability under flux & 99 \%  \\
        \hline
    \end{tabular}
    \end{minipage}
    $^1$ {The readout noise presented here is computed from a CDS noise, and has been measured after correction by the reference pixels (see \citep{NouvelDeLaFleche2022}. Without correction, we find a readout noise of 55e-.}\\
    $^2$ {The dynamic presented here is the full dynamic of the diode, measured with CAGIRE configuration polarisation of the diode, $\Delta V_{diode}$=0.6V. }\\
    \caption{Table of the main characteristics of the detector}
    \label{tab:test_IRFU}
\end{table}

\subsection{Calibration of the detection chain at CPPM}

These tests aim at measuring the behaviour of the CAGIRE detection chain (detector, pre-amplifier and NGC) under CAGIRE conditions (bias voltages, readout mode, signal level...). 
They will also allow to obtain the calibration maps (bad pixels, saturation level, linearity...) for the preprocessing pipeline. 
They are divided into three steps. 

The first step involves the choice of an optimal configuration in terms of bias voltages and number of resets at the beginning of a ramp.

In a second step various tests in darkness will be conducted, with the measurement of the readout noise and of the correlated double sampling (CDS) noise, the identification of hot or erratic pixels and the dark current map. 
Hot pixels are identified by their abnormally high signal in the dark current map, while erratic pixels have a signal abnormally dispersed.

The third step relies on the uniform illumination of the detector with LEDs at two different wavelengths, close to the J and H bands centre (1250\,nm and 1630\,nm). This step will allow to measure the saturation level, the Photo Response Non-Uniformity (PRNU), the ramp non-linearity and the conversion gain on "superpixels", but also to identify cold pixels and erratic pixels. 
Cold pixels are defined by their abnormally low signal under illumination. 
The detector will be illuminated with different levels of flux, corresponding to a signal varying between 80~e-/s/pix to 10\,000~e-/s/pix, in order to measure the flux non-linearity and the impact of persistence. 

The persistence will be studied by alternating illuminations at low flux, close to the expected the sky background ($\sim$160 e-/s/pix in J), and illuminations at large flux, from 250 e-/s/pix to 50\,000 e-/s/pix, with the aim of measuring the amplitude and decay time of the persistent signal, for various excitation levels.

At the end of these tests, the detector will be delivered to IRAP, where it will be integrated into the cryostat.

\subsection{Calibration maps for simulations and the preprocessing pipeline } \label{sub:maps}

Some of the characteristics measured by CEA-IRFU are needed to build realistic simulations of the instrument (section \ref{sec:simu}) and to construct the preprocessing pipeline (section \ref{sec:preproc}). Theses characteristics will be measured at CPPM with CAGIRE configuration before to be implemented on the pipeline. They are presented below.

\subsubsection{Bias and saturation maps}
\begin{figure}[h]
\subfigure{
\centering
\resizebox{0.50\textwidth}{!}{\includegraphics{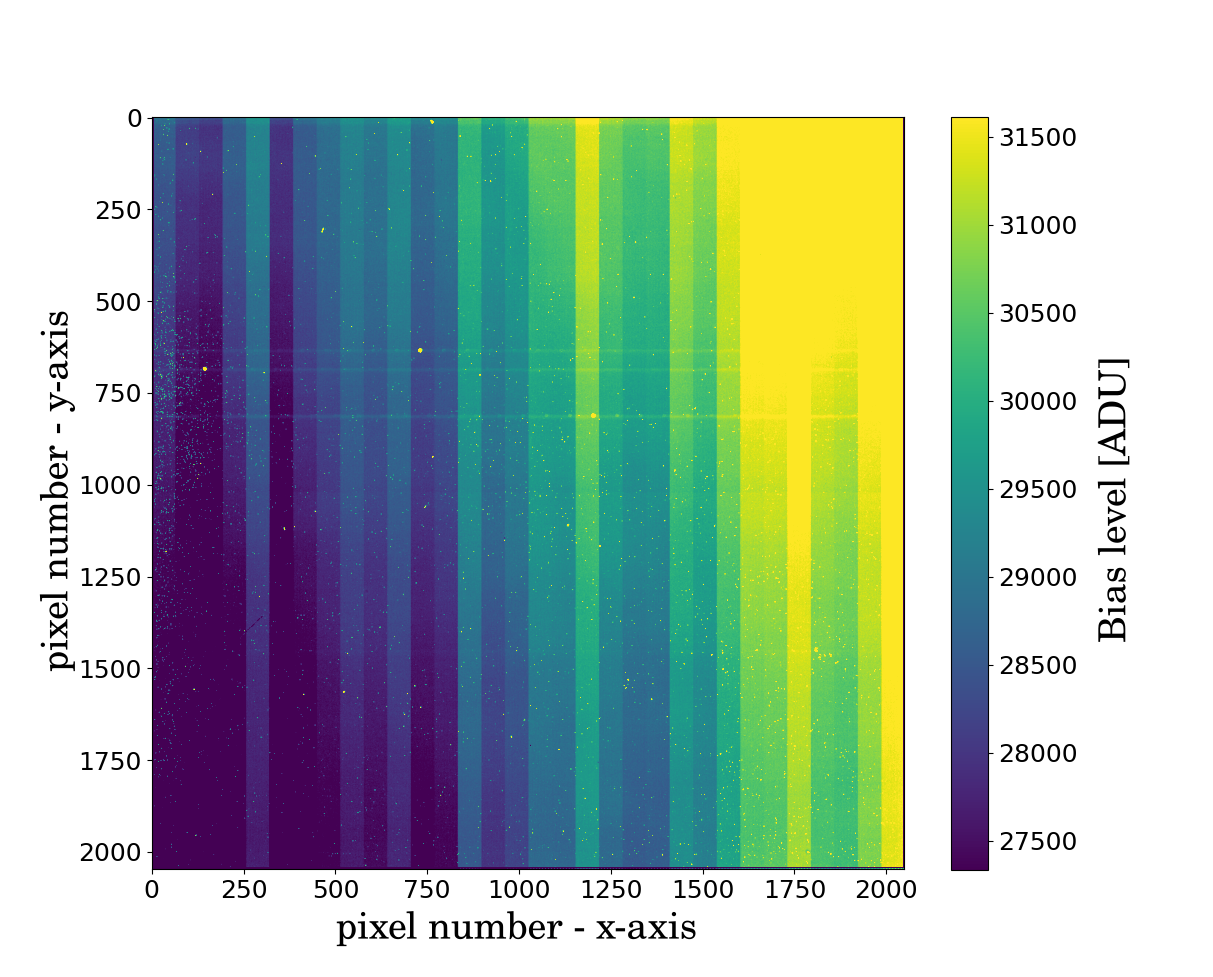}}
}
\subfigure{
\centering
\resizebox{0.50\textwidth}{!}{\includegraphics{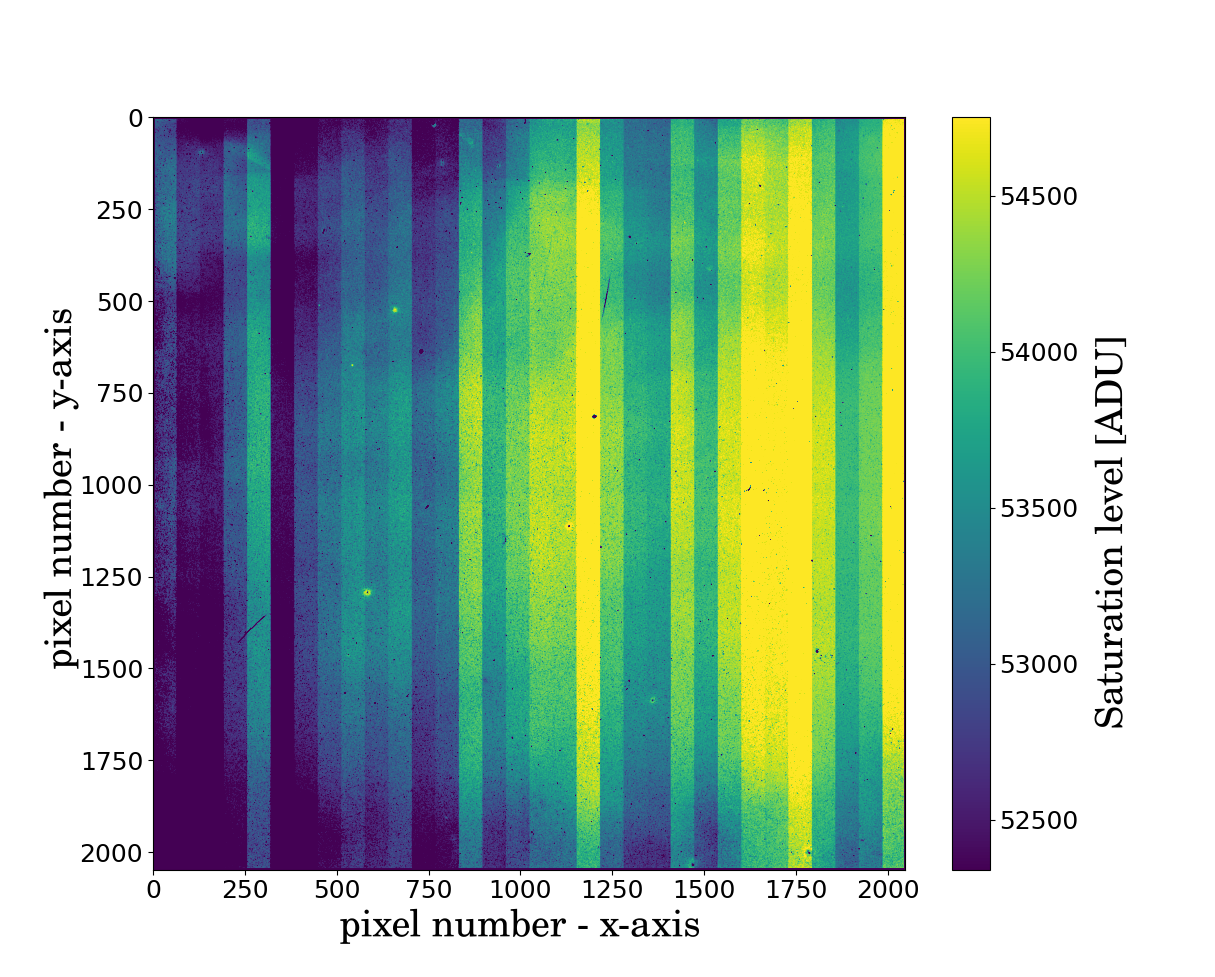}}
}
\caption{a. Bias map (left) and saturation map (right) in ADU.}
\label{fig:saturation_level}
\end{figure}

 The bias calibration map will be measured during characterisations at CPPM.
Figure \ref{fig:saturation_level}a presents a preliminary bias map constructed from measurements done at CEA, which shows that the bias typically varies from $\sim 27$ to $\sim 31$ kADU. 
 
 The saturation level is the signal reached when the potential well of the pixel is full. It is measured by illuminating the sensor until it reaches a constant signal (figure \ref{fig:dk}a). A map of this saturation level is given in figure \ref{fig:saturation_level}b. With a typical bias of $\sim$ 29\,000 ADU and a dynamic of $\sim$ 23\,500 ADU (cf. section \ref{sec:alfa}), the saturation is reached around a median value of 52\,500 ADU.
    
\subsubsection{Non-linearity}
\begin{figure}[h]
\subfigure{
\centering
\resizebox{0.50\textwidth}{!}{\includegraphics{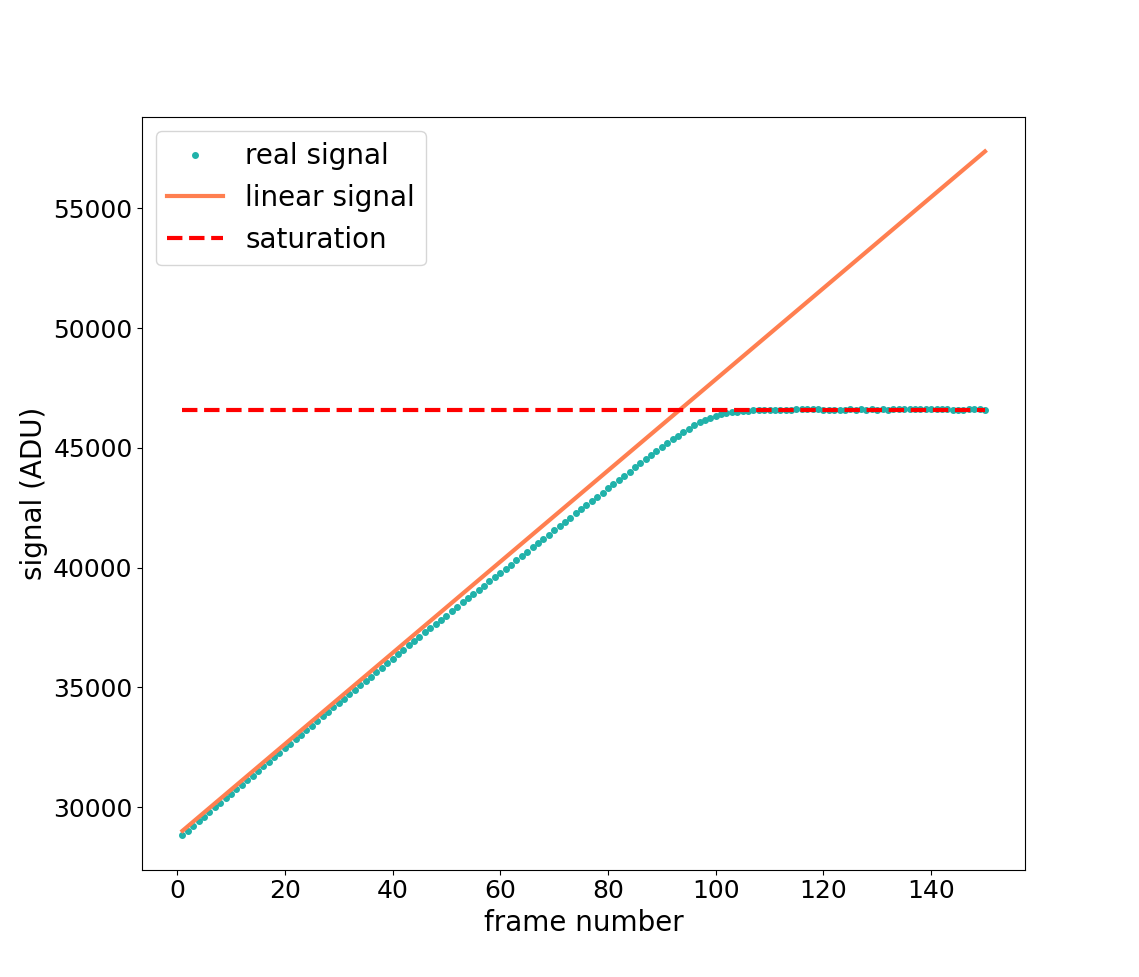}}
}
\subfigure{
\centering
\resizebox{0.50\textwidth}{!}{\includegraphics{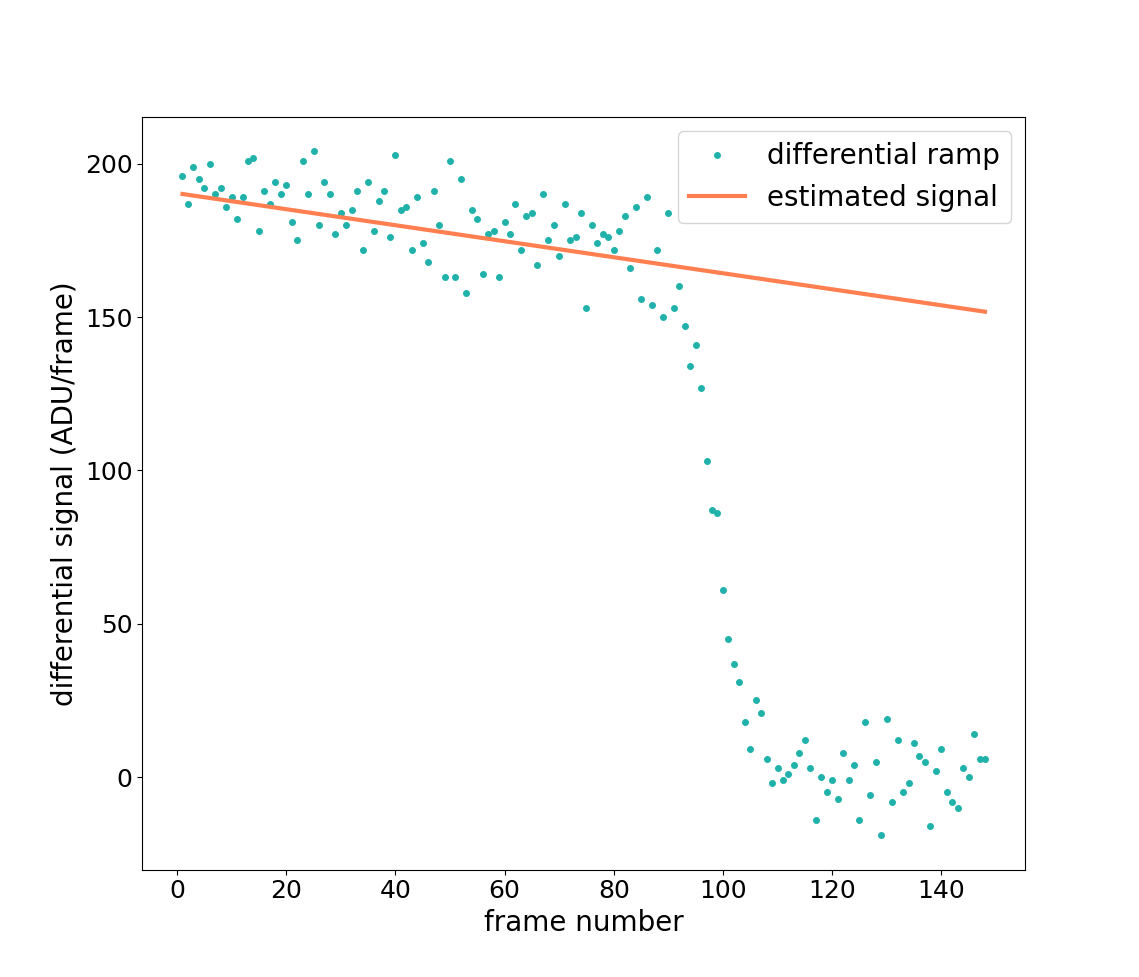}}
}
\caption{a. Experimental ramp (blue point), linear extrapolation of the beginning of the ramp (orange) and saturation level (dashed red line) (left) and b. Differential ramp (blue points) and its linear fit (orange line) (right).}
\label{fig:dk}
\end{figure}

The ramp non-linearity measures the deviation of a \textit{differential ramp}\footnote{The differential ramp is the ramp made by the succession of the differences between two consecutive frames.} $d_k$ measured under constant illumination, from a constant (fig. \ref{fig:dk}). 
Fitting the differential ramp with a linear function, $d_k = A_0 + A_1\times k$, where k is the frame number, yields two coefficients: $A_0$ and $A_1$ respectively representing the offset and the negative slope. This fit is done on the first 70\% of the pixel dynamic range. This threshold allows to fit the ramp in its most linear part, where the deviation from linearity is typically below 5\%. The measurement of the deviation from linearity and the estimation of this range is describe in \cite{NouvelDeLaFleche2022}.  This fit is illustrated in figure \ref{fig:dk}b.

Following \cite{NouvelDeLaFleche2022}, we define the non-linearity coefficient $\gamma$ as: 
    
    \begin{equation}
        \gamma = \frac{A_1}{A_0^2}
    \label{eq:gamma}
    \end{equation}

\begin{figure}[h]
\subfigure{
\centering
\resizebox{0.50\textwidth}{!}{\includegraphics{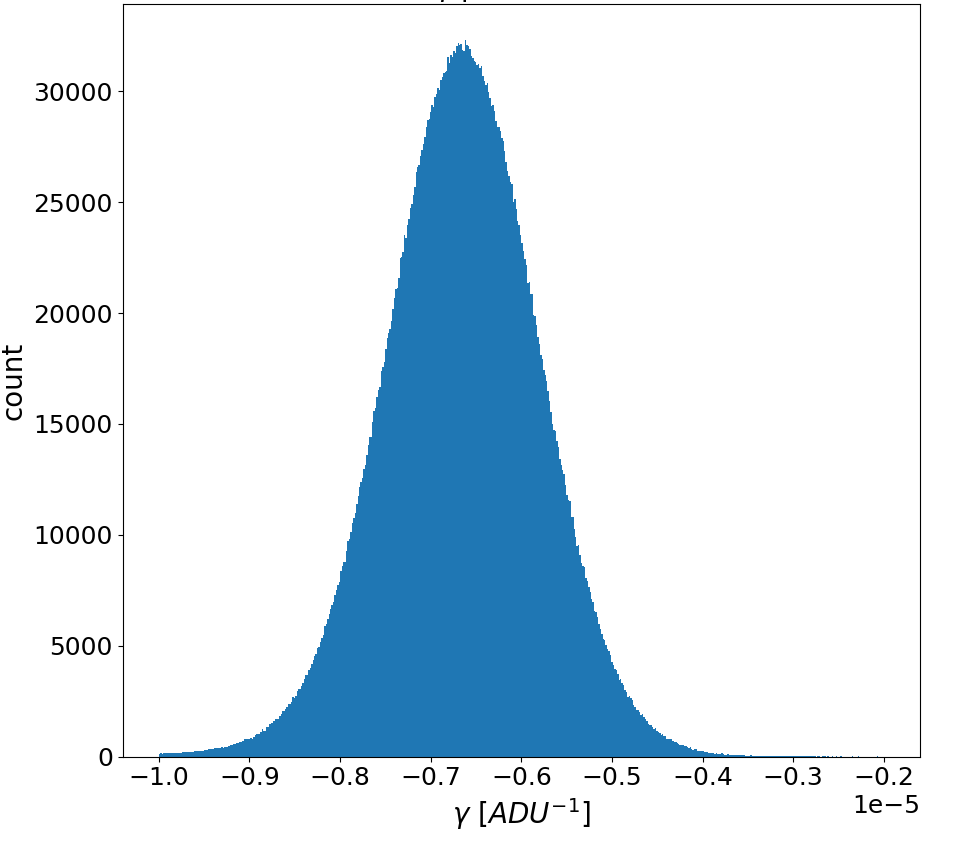}}
}
\subfigure{
\centering
\resizebox{0.50\textwidth}{!}{\includegraphics{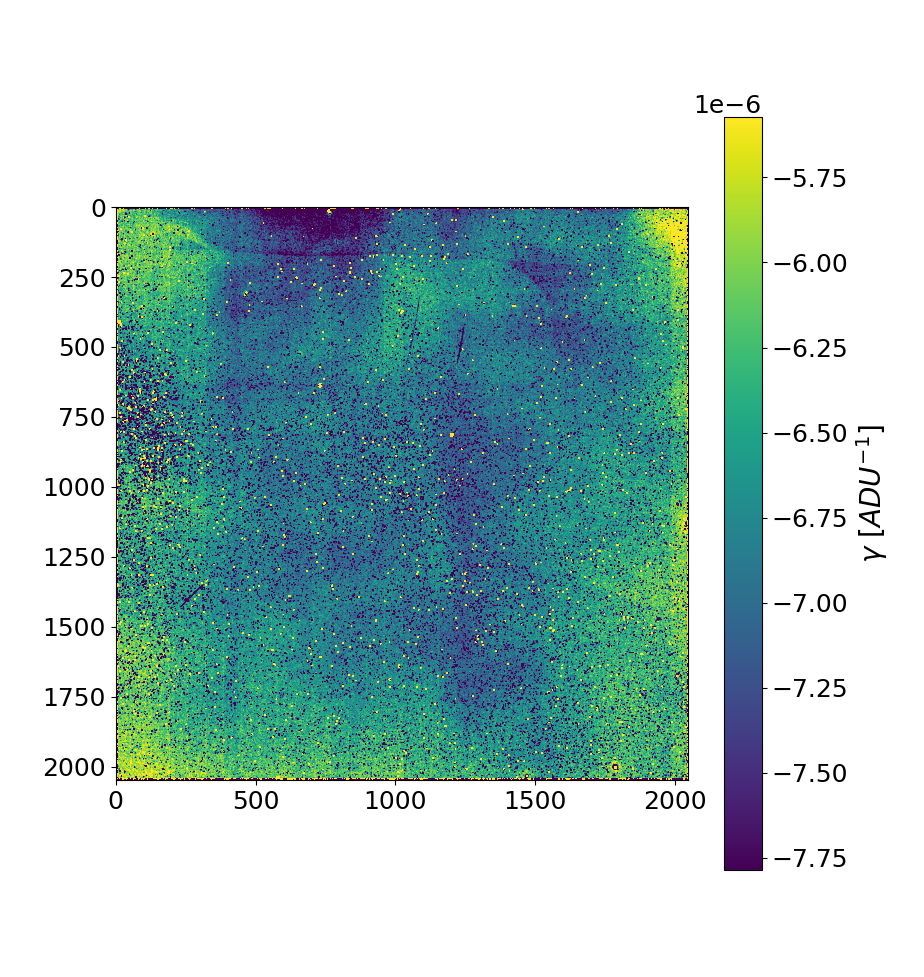}}
%\label{fig:gamma_map}
}
\caption{a. Histogram of the non-linearity coefficient, $\gamma$ ,defined in section \ref{sub:maps} (left) ;  b. map of the non-linearity coefficient $\gamma$ (right) }
\label{fig:gamma}
\end{figure}

This measure of non-linearity is a small negative quantity, which is independent of the signal and has units of ADU$^{-1}$. $\gamma$ encodes the relative loss of signal from one frame to the next under constant flux illumination, and (${1-\gamma}$) may be interpreted as the relative increase of flux that is needed to produce the same signal from one frame to the next.
From the measurements done, the median value of the non-linearity coefficient defined this way is $\gamma = - 6.5 \times 10^{-6}$. An histogram of this coefficient is given in  panel a of figure \ref{fig:gamma} and a map in panel b. A more detailed procedure of the signal measurement with the use of the non-linearity coefficient is given in section \ref{sec:preproc} and in \cite{NouvelDeLaFleche2022}.

\subsubsection{persistent signal}

\begin{figure}[h]
    \centering
    \includegraphics[width=\columnwidth]{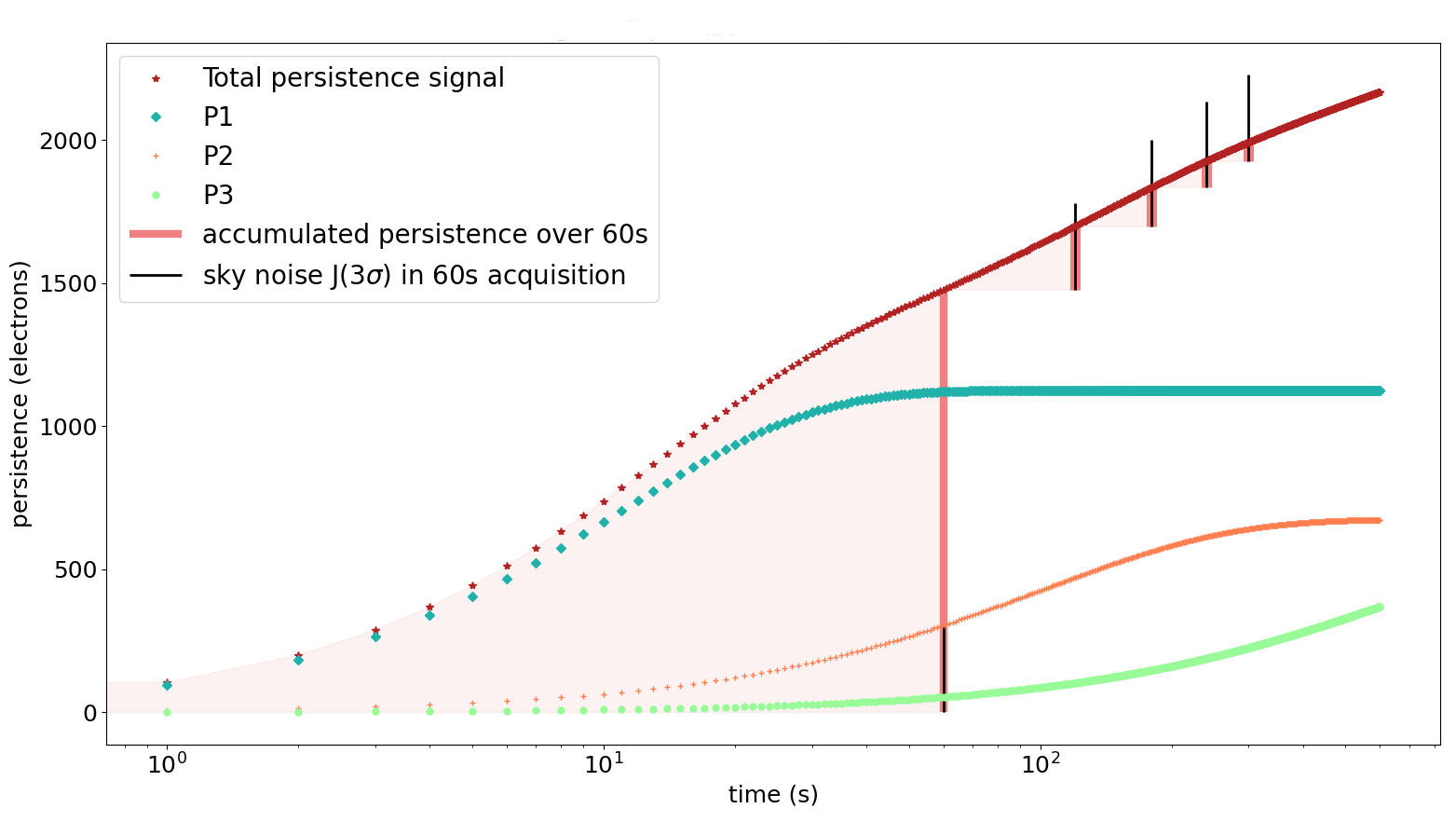}
    \caption{Integrated persistent signal versus log(time), after an illumination at twice the saturation level, at t = 0s. The total signal in electrons, P(t) is plotted with red stars. The blue diamonds, the orange crosses and the green triangles are respectively the first, the second and the third components of equation \ref{eq:persistance}, P1, P2 and P3.
    The pink vertical lines represent the persistent signal accumulated during one 60\,s acquisition, ending at times  t = 60\,s, t = 120\,s, t = 180\,s, t = 240\,s and t = 300\,s. This height must be compared to the dark lines representing the sky noise (3$\sigma$) in J channel for a 60\,s exposure, allowing to evaluate the significance of the persistent signal.}
    %The height of the pink vertical lines represent the persistent signal accumulated during an acquisition of 60\,s, finishing at the pink vertical line. 
    %The height of dark lines represent the sky noise (3$\sigma$) in J channel for a 60\,s exposure, allowing to evaluate the significance of the persistent signal}
    \label{fig:persistance_signal}
\end{figure}

The persistent signal is the signal measured in darkness, in excess of the dark current, after a controlled illumination and various resets of the sensor. As shown by \cite{Legoff2020}, P(t), the integrated persistent signal at time t can be fitted  with a sum of 3 exponentials, as in equation \ref{eq:persistance}. A graph of the integrated signal P(t) for an illumination at 200\% of the saturation level is given figure \ref{fig:persistance_signal}.
    
\begin{equation*}
    P(t) = A_1 \left[ 1- exp \left( -\frac{t}{\tau_1}\right)\right] + A_2\left[ 1- exp \left( -\frac{t}{\tau_2}\right)\right]+ A_3\left[ 1- exp \left( -\frac{t}{\tau_3}\right)\right]
\end{equation*} 

\begin{equation} 
    P(t) = P_1 + P_2 + P_3
\label{eq:persistance}
\end{equation} 

\begin{figure}[h]
\subfigure{
\centering
\resizebox{0.5\textwidth}{!}{\includegraphics{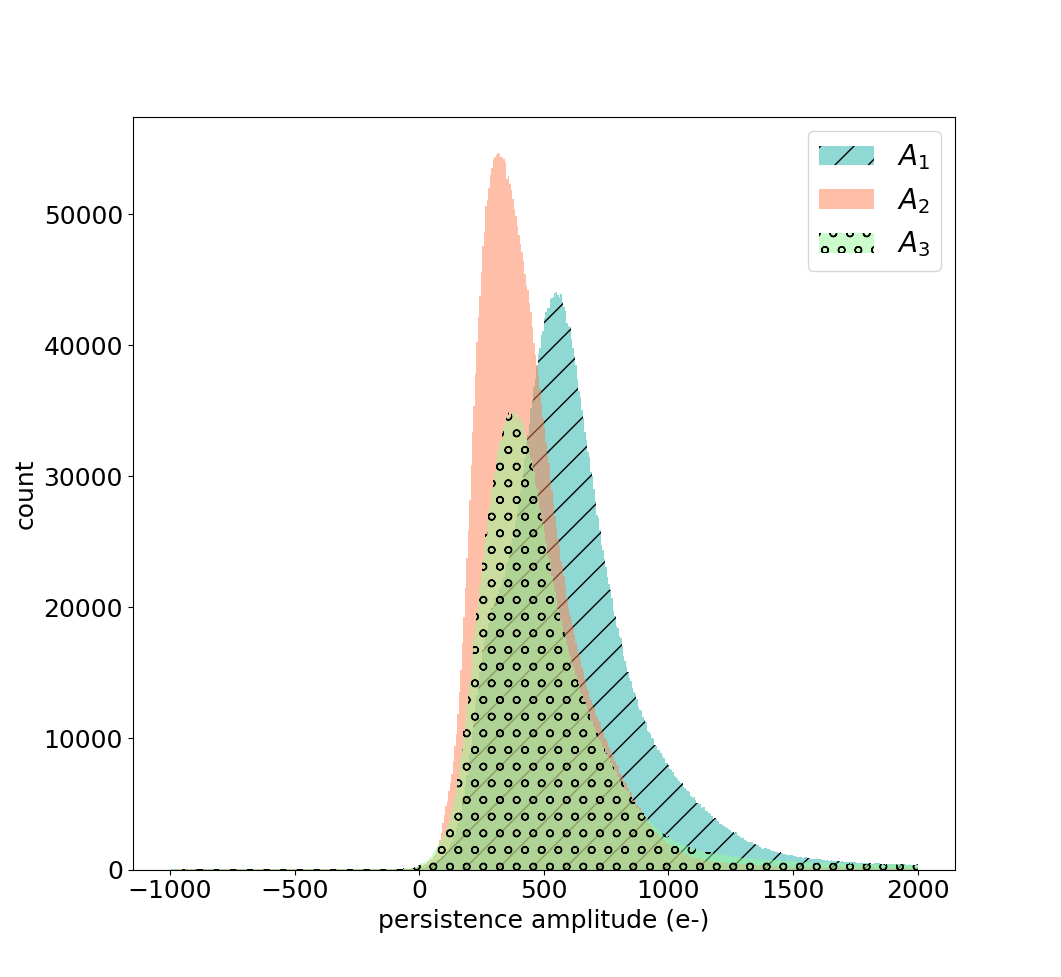}}
\label{fig:amplitude}
}
\subfigure{
\centering
\resizebox{0.5\textwidth}{!}{\includegraphics{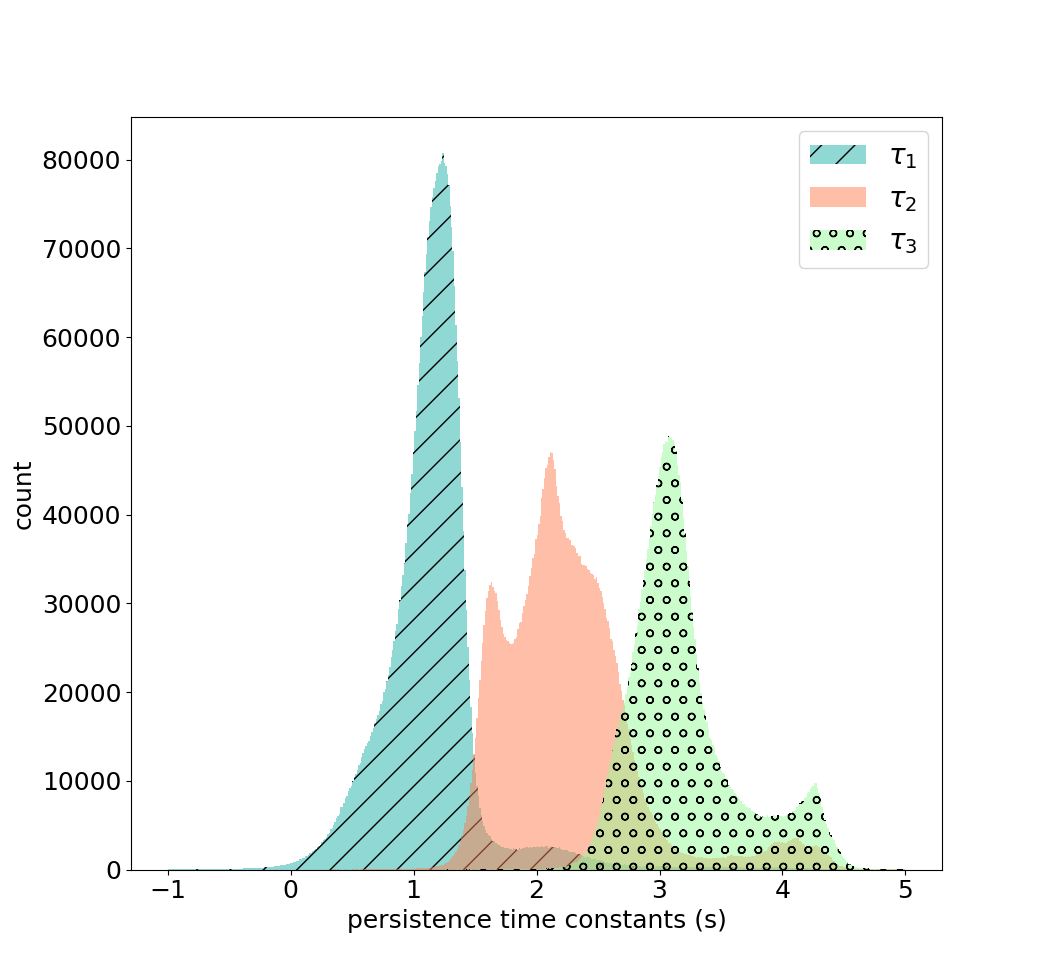}}
\label{fig:time}
}
\caption{ a. Histogram of the amplitude coefficients A$_1$, A$_2$, A$_3$ of persistence defined in equation \ref{eq:persistance} (left) and b. Histogram of the time constants of persistence (in logarithmic scale) defined in equation \ref{eq:persistance} (right)}
\label{fig:pparameters}
\end{figure}

The 6 parameters vary from one pixel to the next and their maps have been computed for different illuminations levels, during tests at CEA-IRFU. 
Moreover, the amplitude parameters $A_1$, $A_2$ and $A_3$ depend on the illumination level.
The distributions of the amplitudes and time constants of the persistent signal are shown in figure \ref{fig:pparameters}, for an illumination at twice the saturation level, suited for our simulations purpose.

The impact of persistence is illustrated in figure \ref{fig:persistance_signal} showing the evolution of the integrated persistent signal after an illumination at twice the saturation level, at t = 0. 
The pink vertical segments represent the persistent signal accumulated during a 60\,s long exposure taken at various times after the illumination (from 0 to 4 minutes). 
The persistent signal can be compared with the sky fluctuations, which amount to $\sim 300$ e- for a 60\,s long exposure in J band ($3\sigma$).
Figure \ref{fig:persistance_signal} shows that, only for the ramp immediately following the illumination, the persistent signal dominates over the sky noise, and must be taken into account.
Conversely, an illumination at t = 0 will not strongly affect exposures taken more than 1 minute after the illumination. 

\subsection{Calibration of the camera at IRAP}

After the characterisation of the detection chain at CPPM, the whole camera will be integrated and characterised at IRAP. 

\subsubsection{Anticipated tests}
Before carrying out the integration of the camera, several preliminary validations are performed, to validate each block of the camera:
\begin{itemize}
    \item Measurement of the conversion gain of the NGC alone and of the NGC linked to the pre-amplifier.
    \item Check of the temperature range supported by the NGC, and adaption of the rack to make it work between -15°C and 35°C (external temperature at the San Pedro M\'artir observatory). 
    \item Adaptation of the NGC configuration files to CAGIRE needs.
    \item Test of the readout chain (NGC+PA+ROIC) on the telescope, to verify whether electromagnetic interferences impact the readout chain. This test was made possible thanks to the presence of the telescope at ``Observatoire de Haute Provence'' (OHP) for calibration. No specific noise resulting from the telescope motorisation was detected during this test. 
    \item Validation of the cryocooler performance over long periods of time (several months).
    \item Unitary controls of optics quality (dimension, curvature, coating).
    \item Development of a star projector to simulate point sources on the detector and to validate the alignment procedure. 
\end{itemize}

After the reception of the detector and the cryostat at IRAP, the whole camera will be integrated and characterised. \\

\subsubsection{Camera integration and validation}
The validation of the whole camera involves various activities and measurements. Crucial activities include the integration of the detector into the cryostat, the validation of the cryocooler performance (the capacity to reach the required temperature at the detector), the verification of the vacuum tightness of the cryostat and the validation of the alignment procedure.

Following the instrument integration, several measurements will permit the detailed characterisation of various features of the camera. The list includes: the dark level measured with the internal cold shutter; the flatfield with the help of an external blackbody placed in front of the camera; the thermal emission inside the cryostat with an external cold screen in front of the camera; the response to point sources with a specially designed star projector; the impact of the J and H filters on the camera focus; and the verification of the spectral range of the camera with a monochromator. 

The end of the camera Assembly Integration Test / Assembly Integration Validation (AIT/AIV) will be validated by the Delivery Test Review, planned in the first quarter of 2024. 
This review will allow the delivery of the camera at the National Astronomical Observatory in San Pedro Mártir, Mexico (OAN-SPM), where it will be mounted on the telescope.
Before the installation of CAGIRE, the telescope will operate with mass models of the cryostat and of the close electronics attached to the Main Structural Unit to balance the telescope and allow its operation with the two visible cameras.

\section{Image simulation}
\label{sec:simu}

We present here a simulation tool that produces realistic CAGIRE ramps. 
Such simulated ramps have been used to test the preprocessing pipeline, to estimate the performance of the camera and to adapt the observing strategy. 

The simulation is based on an Image Simulator (IS), developed by D.\,Corre \citep{Corre2018}. This IS aims at computing realistic images, taking into account catalogs of the sky, the impact of the atmosphere, the telescope and detector response, and the impact of cosmic rays. 
The computation is based on an Exposure Time Calculator (ETC) also designed by D.\,Corre. 
This ETC provides, among others parameters, the zero-point of CAGIRE at the focus of COLIBRI, calculated from the telescope characteristics, the site parameters and the detector characteristics. 

The Image Simulator has been adapted with the real characteristics of CAGIRE, by adapting the telescope and detector input files with the characteristics of CAGIRE. 
One significant adaptation was the production of ramps representative of the "Up-The-Ramp" operation mode of CAGIRE.
We also customised the sky, environment and detector description as described below. 

\begin{figure}[]
    \centering
    \includegraphics[width=\columnwidth]{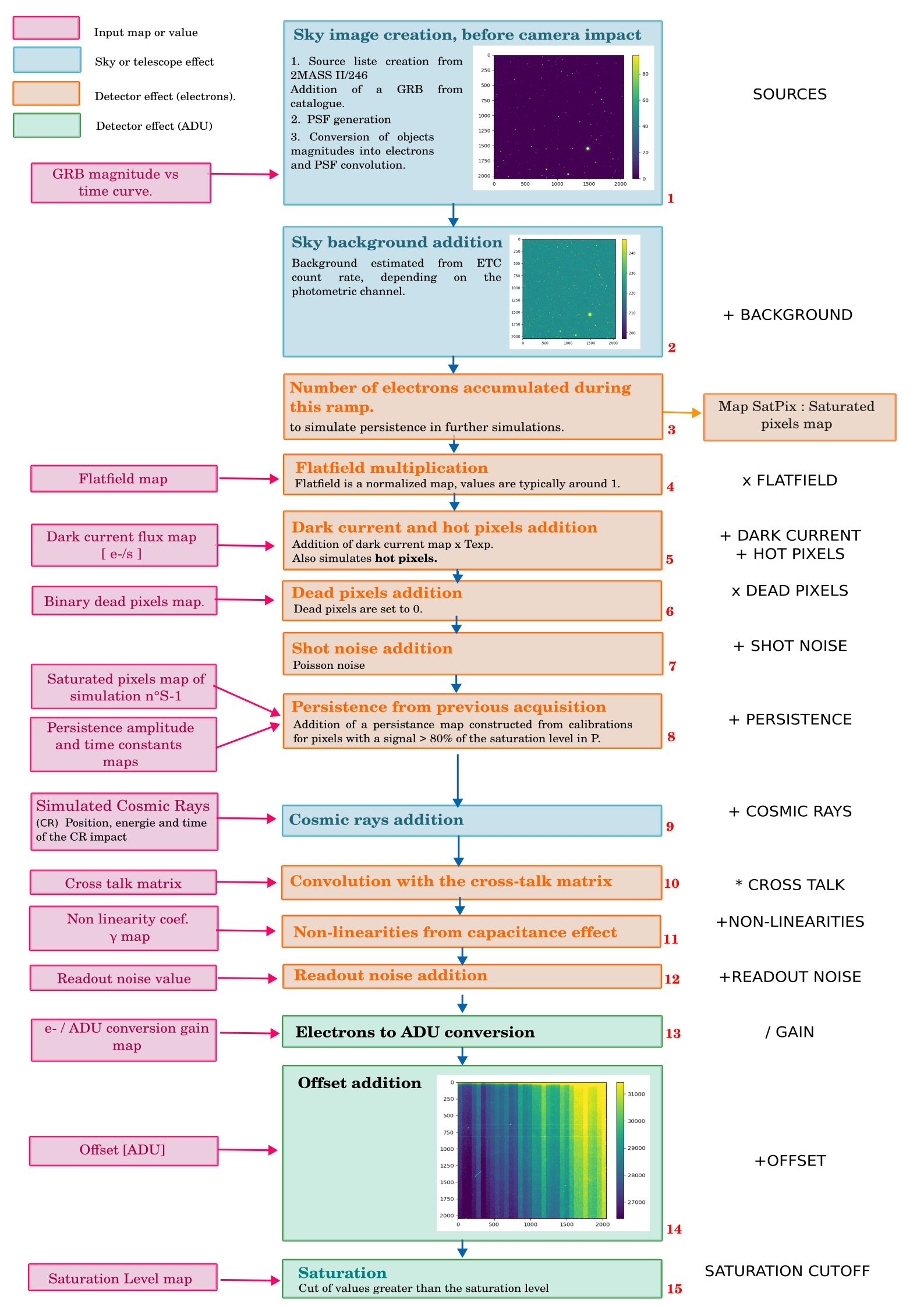}
    \caption{Steps of the simulation}
    \label{fig:simuDiagram}
\end{figure}

The NIR sky considered in the simulator uses the sky background computed by the ETC and stars from the \textit{2MASS} catalog \citep{Cutri2003}. 
A notable evolution is the possibility to add a decaying GRB, whose signal is computed for each frame of the ramps,  while simulating the "Up-The-Ramp" mode.
The description of the environment is limited to the addition of cosmic rays, whose impact is modelled by a number of additional counts in randomly chosen pixels. 
Considering the detector, we take into account the following effects: the dark current, the (non-uniform) response to photons measured through QE measurements, the detector's cosmetic (dead pixels and hot pixels identified during calibrations at CPPM), the interpixel cross-talk (measured by CEA-IRFU), the ramp and flux non-linearities, the persistence from the previous acquisition and the readout noise. The flux non-linearity will not be detailed here as its calibration has still to be  conducted. 
The simulation finally adds the photon statistics, converts electrons to ADU (Arbitrary Data Unit), adds the bias, and takes into account the saturation level of each pixel. 
The simulation of some of these effects uses calibration maps measured during the characterisation of CAGIRE, which have been described in the previous section. 
A diagram of the steps involved in the simulation of one frame of the ramp is given fig. \ref{fig:simuDiagram}, and described in \ref{sub:simupar}. 

The overall process depends on the exposure time T$_\mathrm{exp}$ of each frame, which is a multiple of 1.33\,s. To compute a full ramp, the simulation of one frame is repeated, with T$_\mathrm{exp}$ increased by 1.33\,s at each frame. The succession of these frames makes the ramp.

\subsection{Simulation parameters}
\label{sub:simupar}

We will describe here the succession of the different steps of a simulation. Before to launch a simulation, the user can choose some parameters, listed hereafter : 

\begin{itemize}
    \item The photometric channel studied bands = ['J' or 'H'].
    \item The sky region to simulate with the central coordinates: RA$_{image}$=[degrees]; DEC$_{image}$=[degrees]. 
    \item The exposure time of the total ramp, in seconds.
    \item If the user want to simulate a GRB, he needs to specify the GRB to simulate to import the information from the GRB database we have created. This database contain the magnitude of the GRB versus time, and its position.   
    \item If a GRB is simulated, the image will be centred around a random position, at a maximum of 0.16° from the simulated GRB, to be sure to simulate the GRB inside CAGIRE field of view. 
    \item The time between the burst detection and the start of the observation by the telescope, in seconds.
\end{itemize}

Then, the simulation can be launched. The different steps of the simulation are identified in figure \ref{fig:simuDiagram}, by a small number at the bottom right of each box.
The input parameters (or maps) of the simulation are shown inside pink boxes in figure \ref{fig:simuDiagram}. Most of them have been measured during the detector calibration. 
The central boxes represent the different steps of the simulation, regarding the telescope effect (blue) the detector effects in electrons (orange) and the detector effects in ADU (green). 
The main goal/effect of each step is summarised in the list of words at the right of the diagram. 

We will now detail more specifically some steps mentioned in the diagram : the GRB simulation, the Non-linearity simulation and the persistence simulation.

\subsection{GRB simulation}
\label{sub:grbsim}

The first step of the simulation is the computation of the signal of simulated sources at each frame. 
For constant sources, a constant signal is added. 
For GRBs, their magnitude is estimated at the time of each frame and the corresponding signal is added, allowing varying light-curves. 

In order to test the sensitivity of CAGIRE to high-z bursts, we have created a database of several GRBs with a redshift greater than z=5, detected in the J and H channels over the past few years. We compute their light curves from the first detection up to 1 day after the first detection. Thanks to this database, when a GRB is added to a simulated map, the signal of the GRB is updated at each frame.

\subsection{Simulation of ramp non-linearities}
\label{sub:nonlin}

The simulation of ramp non-linearities is based on the coefficient of non-linearity, $\gamma$, which measures the linear deviation of a \textit{differential ramp} $d_k$ measured under constant illumination from a constant signal. Using this coefficient $\gamma$ described in section \ref{sub:maps}, the signal $S_k$ accumulated at frame k of a simulated ramp can be expressed by equation \ref{eq:signal}, with $a_0$ the slope of the simulated ramp. 

\begin{equation}
    S_k = a_0k +  a_0^2 \times \gamma \times \frac{k \times (k+1)}{2}
\label{eq:signal}
\end{equation}

For a linear ramp, $S_k$ would be equal to the first term $a_0k$.  The second term of equation \ref{eq:signal} is negative, it represents the departure from linearity due to the ramp non-linearity. 

\subsection{Simulation of persistence}
\label{sub:persistence}

The simulation only deals with persistence caused by previous illuminations and does not address the persistence due to the ongoing illumination, which is taken into account by our modelling of the flux non-linearity. We describe here steps 3 and 8 of the diagram in fig. \ref{fig:simuDiagram}.

While most of the simulated effects are well known, persistence remains difficult to calibrate and thus to simulate. 
Our simulations rely on two elements: the identification of saturated pixels in previous exposures and the application of a model describing how the persistent signal evolves with time.
This model, which has been constructed by CEA-LETI (see section \ref{sub:maps}), describes the persistent signal $P(t)$ with a sum of 3 exponentials (eq. \ref{eq:persistance}), with typical decay times of 15 - 150 - 1500\,s and amplitudes adjusted to reproduce the observed persistent signal of each individual pixel (measured in darkness after a controlled illumination, \cite{Legoff2020}).
While this model depends of the full history of previous illuminations, we consider here two simplifications: the persistent signal is computed for an illumination at two times the saturation level and only for pixels saturated in the immediately preceding exposure.
The first simplification is justified by the measurements done at CEA, which have shown that above two times the saturation level, the amplitude of the saturation does not change much.
The second simplification is justified by the rapid decrease of the persistent signal with time, which reduces the impact of persistence when the saturation occurs before the previous ramp.
This is illustrated in figure \ref{fig:persistance_signal}, which shows that after 1 minute ($\sim$ one ramp) the persistent signal is below the sky fluctuations.
Saturated pixels are identified during the immediately preceding simulation, as shown in step 3 of the diagram in fig. \ref{fig:simuDiagram}. 
Knowing the saturated pixels, their persistent signal can be computed for each frame and added to the simulated ramp.

For a given pixel, the persistent signal depends on the time elapsed since the end of the previous ramp, of the ramp duration, and on the 6 coefficients describing the 3 exponentials used to model the persistence. \\

To conclude, the simulations are important tools to understand the impact of detector effects in CAGIRE images, but also to test the pipeline presented in section \ref{sec:preproc}. The time needed to complete a simulation depends mainly on the length of the ramp and of the number of sources simulated. Typically, simulating a 60\,s ramp on a sky region with 200 sources takes $\sim$10 min on a PC. The simulations will continue to evolve, and become more reliable with the calibrations to come at CPPM.

 \section{Preprocessing pipeline}
\label{sec:preproc}

One goal of COLIBRI is the fast identification and accurate localisation of GRB afterglows, within a delay of 5 minutes. 
This requirement has strong implications on the operation of the telescope in general and CAGIRE in particular.
One of them is the need to provide clean images to the astronomy pipeline, quickly after the acquisition of a ramp.

This is the role of the preprocessing pipeline (hereafter Preproc), which converts raw ramps into images corrected for various detectors and environmental effects.
The main issue for this pipeline is to quickly process 4 millions of pixels, with their own individual response, before the end of the next acquisition. 

\begin{figure}[]
    \centering
    \includegraphics[width=\columnwidth]{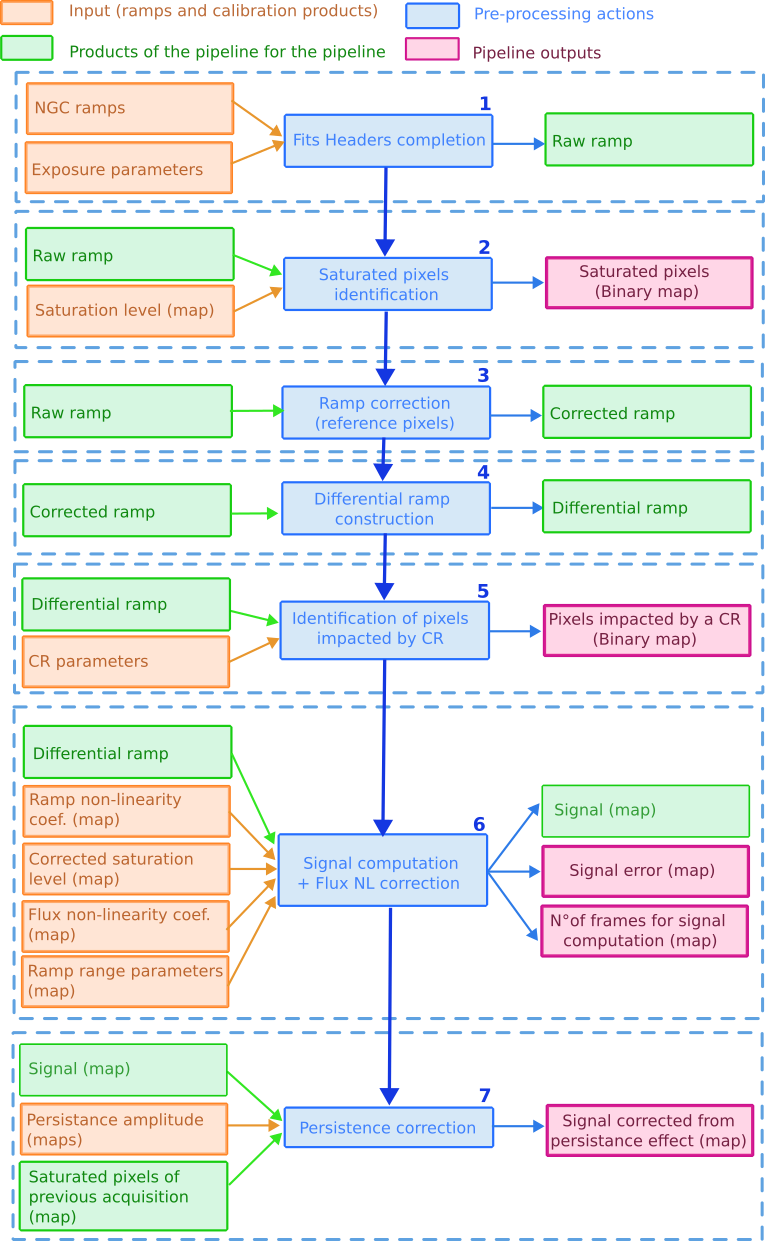}
    \caption{Diagram of the preprocessing pipeline. If Step 7 is not done, the signal map of step 6 is considered as the output }
    \label{fig:preproc}
\end{figure}

The pipeline is composed of 6 main steps illustrated in fig. \ref{fig:preproc}, some of them using pre-existing calibration maps: 

\begin{itemize}
    \item The completion of the fits headers of the NGC ramp, produces the so-called \textit{Raw ramps}. 
    
    \item The identification of saturated pixels, with the help of a ``saturation level'' calibration map. This produces a binary map of saturated pixels. 
    
    \item The correction of each frame of the ramp from common mode noise and offsets thanks to the reference pixels, leading the \textit{corrected ramp}. 
    
    \item The construction of the \textit{differential ramp} by subtracting two consecutive frames of the corrected ramp. 
    
    \item The identification of pixels impacted by cosmic rays and the computation of their signal thanks to their peculiar signature in the differential ramp. This produces a binary map of pixels impacted by cosmic rays. 

    \item The signal computation, based on the slope of the ramp. This computation includes corrections for two types of non-linearities: the ramp non-linearity and the flux non-linearity\footnote{We call \textit{ramp non-linearity}, the deviation of counts in a ramp from a straight line. This deviation is mostly due to the increase of pixel capacitance with the growing number of charges accumulated in a pixel. We call flux non-linearity, the fact that the linear extrapolation of a ramp is not exactly proportional to the flux received by the pixel.}. 
    This step uses four calibration maps: the ramp non-linearity coefficients (section \ref{sub:maps}), the flux non-linearity coefficients, the map of the saturation level and the offset-subtracted saturation level. 
    It produces three outputs maps: the signal, the signal variance, and the number of frames used to compute the signal. This last step is described in more details in \cite{NouvelDeLaFleche2022}.

    \item Persistence correction. As we have seen in section \ref{sub:persistence}, the persistence from a ramp may strongly impact subsequent ramps. To mitigate this effect, we propose to compute, for each ramp, the persistence expected from saturated pixels in the previous ramp. This correction involves two time intervals: $\mathrm{T_{reset}}$, the delay between the end of the previous ramp and the start of the current ramp, and $\mathrm{T_{tot}}$, the delay between the end of the previous ramp and the end of the current ramp. 
    The persistent signal is computed using a simplified version of equation \ref{eq:persistance} with only the two first exponentials (smaller constant time), where $\tau_1$ and $\tau_2$ have fixed values: $\tau_1 = 15$\,s and $\tau_2 = 150$\,s and calibration maps are used to measure A$_1$ and A$_2$. 
    The goal of this simplification is to limit the number of input maps, while keeping sufficient accuracy of the model on the second to minute timescale. 
    In the end, the persistent signal $S_p$ is computed as the difference $\mathrm{S_p = P(T_{tot}) - P(T_{reset}})$ for all active pixels, where P(T) represents the persistent signal integrated up to time T. 
    This signal integrated from $\mathrm{T_{reset}}$ to $\mathrm{T_{tot}}$ is then converted into a signal in e-/s, and 
    subtracted from the signal of the current ramp. 
    The efficiency of this method is discussed in section \ref{sub:persistence}

\end{itemize}

\begin{figure}[h]
    \centering
    \includegraphics[width=\columnwidth]{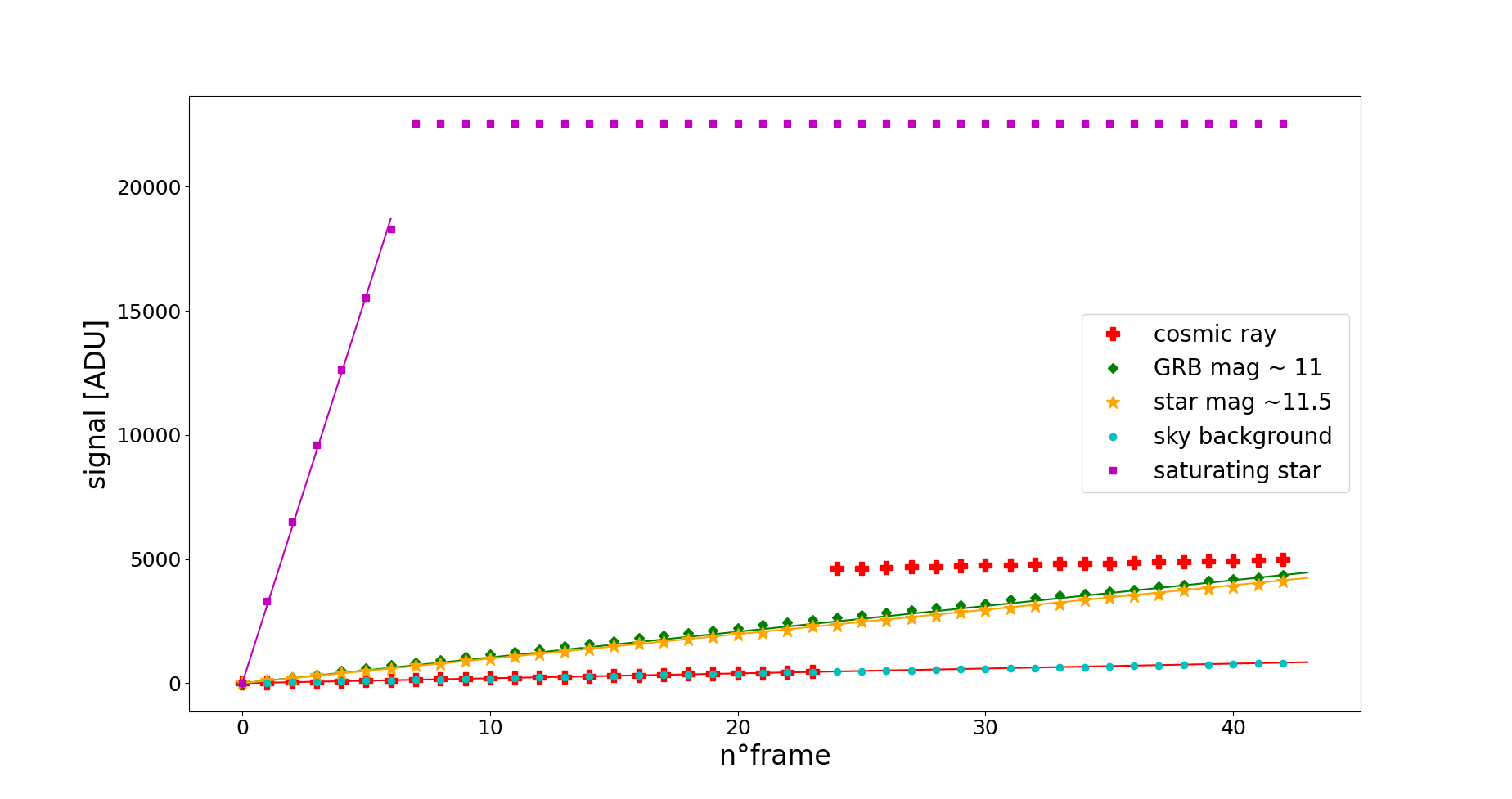}
    \caption{Simulated ramps (dotted lines) and associated signal computed by the pipeline (full lines) for a pixel impacted by a cosmic ray (red crosses), for a simulated GRB (green diamonds), for a star with a magnitude similar to the GRB one (orange stars), for sky background (blue points) and for a saturating pixel (purple square). The signal measured for the sky pixel impacted by a cosmic-ray (red line) is equal to the sky signal (blue line)}
    \label{fig:rampes}
\end{figure}

\begin{figure}[h]
\subfigure{
\centering
\resizebox{0.50\textwidth}{!}{\includegraphics{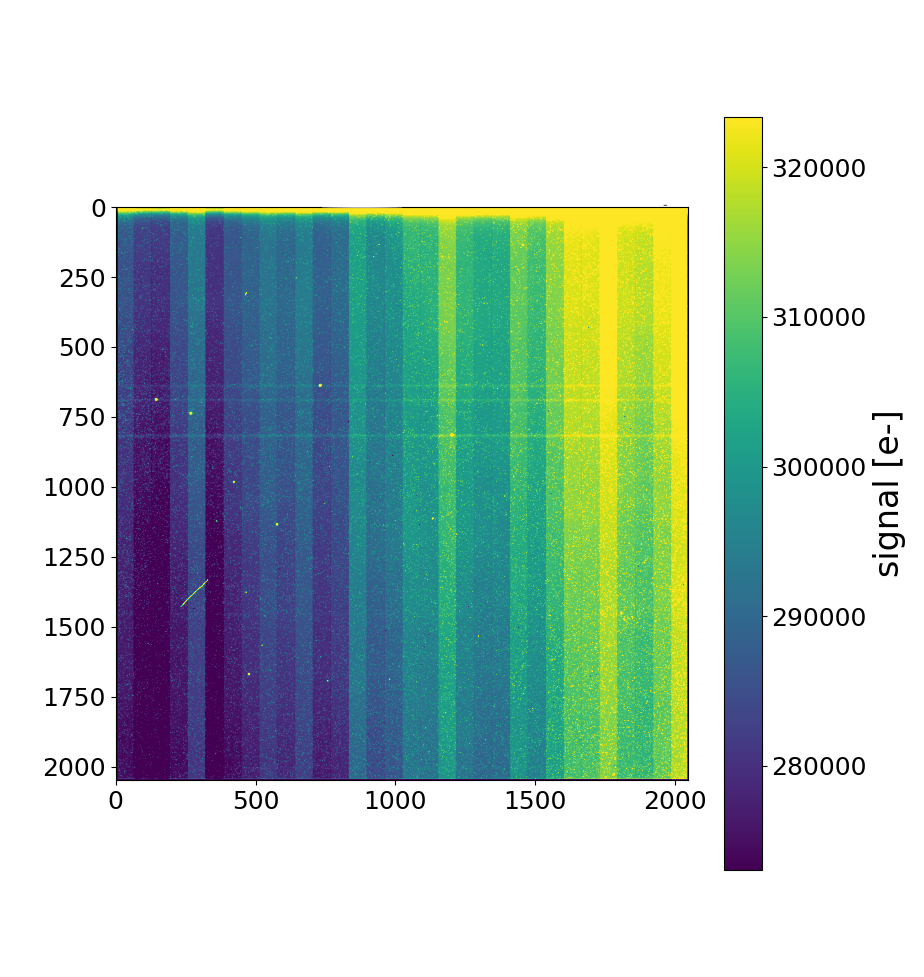}}
\label{fig:last_frame}
}
\subfigure{
\centering
\resizebox{0.50\textwidth}{!}{\includegraphics{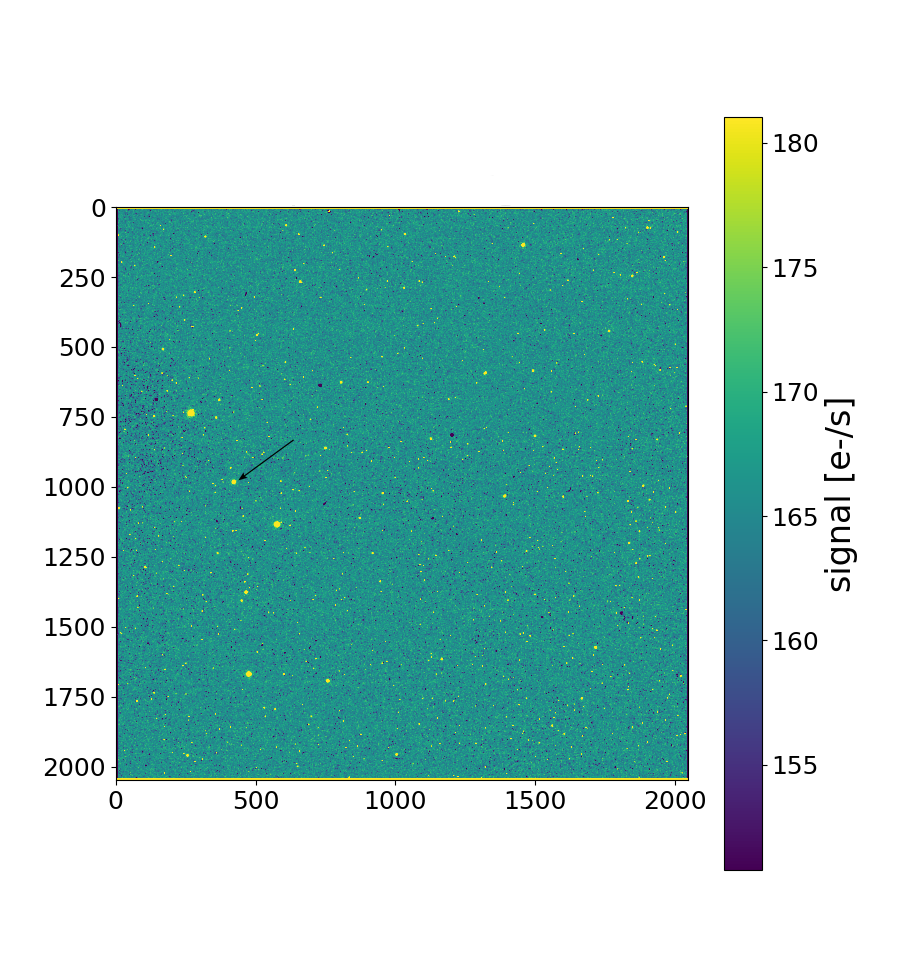}}
\label{fig:preproc_result}
}
\caption{a. Last frame of the ramp (left) and b. Preprocessed Signal ready for the astronomy pipeline (right). The black arrow points the simulated GRB }
\label{fig:presult}
\end{figure}

The result of the Preproc is shown in Figure \ref{fig:rampes} for some particular pixels. 
This figure illustrates the ability of the Preproc to measure the signal of normal pixels (in green, yellow, and blue), of saturated pixels (in purple), and of pixels impacted by a cosmic ray (in red).
The images produced by the preprocessing are recorded as extensions into a single fits file, which is made available to the TCS, while the raw ramps are stored in an archive for future use, if needed.
An example of a preprocessed image is presented in figure \ref{fig:presult}b, next to the last frame of the ramp in figure \ref{fig:presult}a, where the signal has been computed from a simulated ramp of a sky region centered around GRB\,090423 ([RA - DEC]= [148.7834° - 18.167°]).

The validation of the preprocessing has been done with both real and simulated ramps. Real ramps have been recorded with the RATIR instrument\footnote{the Reionization And Transients InfraRed camera.} \citep{Butler2012}, a camera equipped with two H2RG SWIR detectors from the Teledyne company. 
It was possible to use these ramps to test the preprocessing because the ALFA detector is very similar to the H2RG used in the near infra-red channels of RATIR. 
The use of simulated ramps to test the preprocessing is explained in section \ref{sec:pvalid}. 

In terms of performance, we have demonstrated in \cite{NouvelDeLaFleche2022} the ability of the preprocessing to quickly process the ramps, nearly within half the acquisition time of one ramp. 
The preprocessing is independent of the acquisition, as it runs on the ``CAGIRE computer'', which is distinct from the acquisition computer (the LLCU, see section \ref{ssub:acqChaine}). 
In this way, a problem with the preprocessing cannot block the acquisition of future ramps.
More information about the preprocessing and its validation with sky images is given in \cite{NouvelDeLaFleche2022}.

\section{Validation of the preprocessing pipeline}
\label{sec:pvalid}
This section aims to analyse the impact of the corrections performed by the Preproc, such as the use of the non-saturated part of a ramp for signal computation, the non-linearity correction, the flatfield subtraction, and the correction for persistence. 
It relies on a 60\,s long simulated ramp with the J filter, constructed in the field of GRB\,090423, detected by GROND at [RA; DEC] = [148.895° ;  18.6°], with a redshift $z= 8.26$. The simulation starts 5\,min after the GRB.
The corresponding preprocessed image is shown in figure \ref{fig:presult}b, where the position of the GRB is indicated with an arrow.

The verification works as follows: the simulated ramp goes through the Preproc, producing a signal map that is then divided by the flatfield map, and searched for sources with the Source-Extractor software \citep{Bertin1996}. These last two operations crudely simulate the work of the astronomy pipeline. 
The sources extracted are cross-matched with the \textit{2MASS} catalog, which was used as input, and their magnitudes compared with the catalog's ones, allowing to measure the impact of the Preproc.
The goal of the crude analysis presented in this section is not to assess the performance of CAGIRE, but simply to verify that the preprocessing works properly and produces no side effect when applied to realistic simulated ramps. 

\subsection{Overall performance}
\label{sub:overall}

When we compare the sources extracted in our file with those in the \textit{2MASS} catalog, we obtain 97.5\% of good detections and identifications. 
Five sources are found by Source-Extractor which are not referenced in the \textit{2MASS} catalog: they are the simulated GRB and 4 false stars due to the persistence of saturating sources in the previous exposure (with J magnitudes ranging from 9.1 to 11.1)

\begin{figure}[h]
\subfigure{
\centering
\resizebox{\textwidth}{!}{\includegraphics{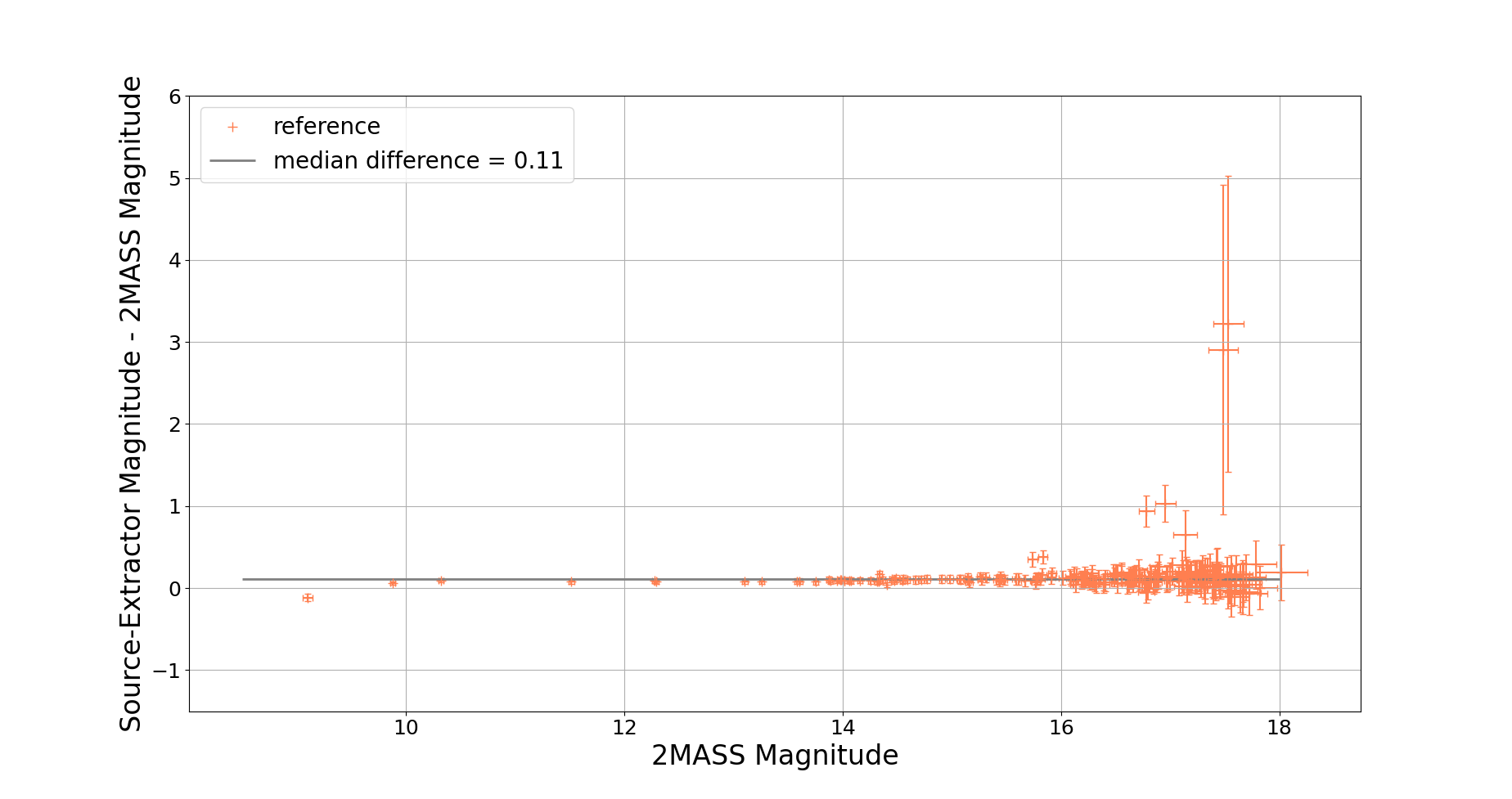}}
\label{fig:2massA}
}
\subfigure{
\centering
\resizebox{\textwidth}{!}{\includegraphics{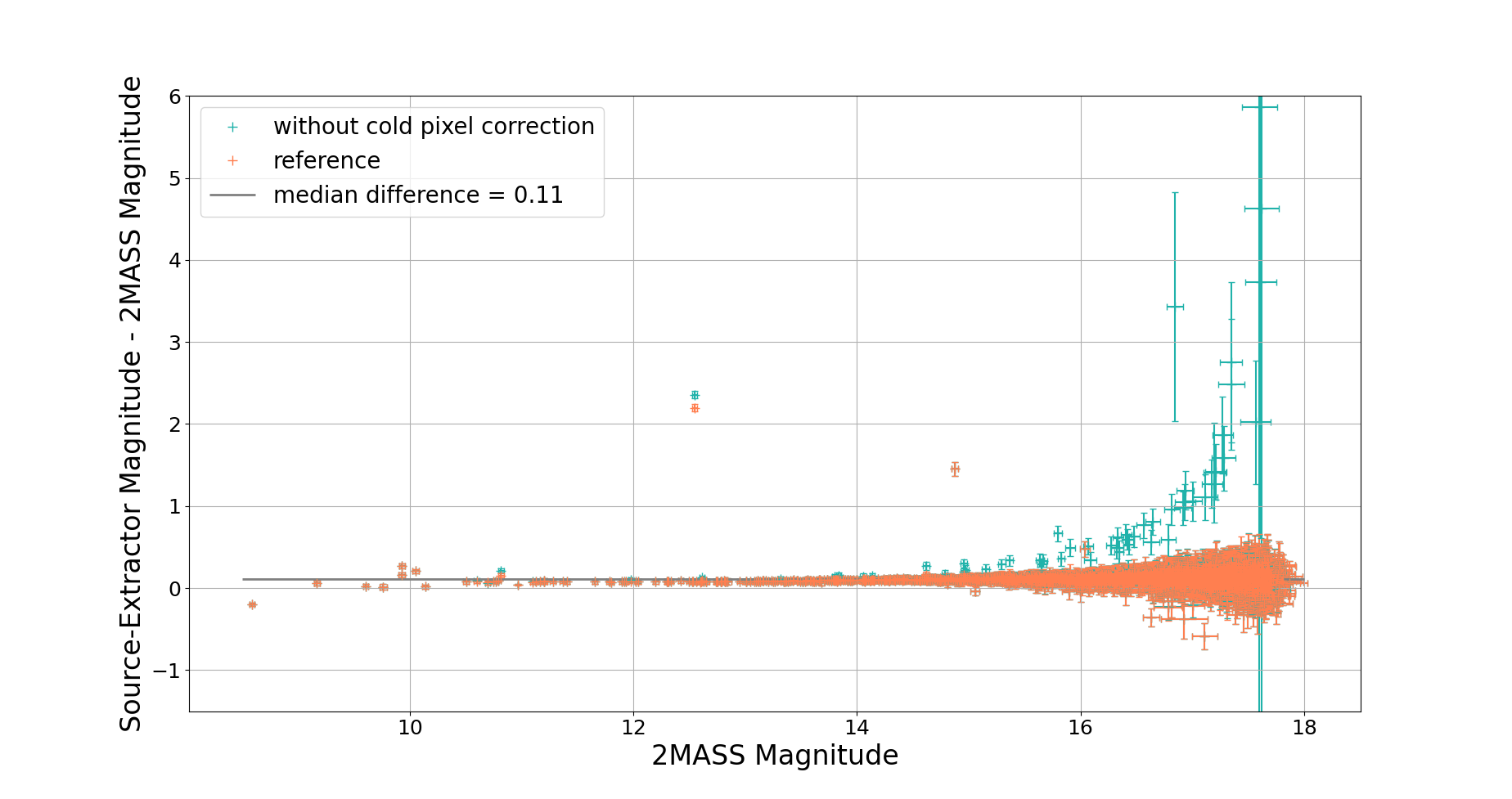}}
\label{fig:2massB}
}
\caption{Difference between magnitudes estimated by Source-Extractor and \textit{2MASS} magnitudes as a function of \textit{2MASS} magnitudes. a. For a field of view centred around [RA; DEC] = [148.895° ;  18.6°] and b. For a field of view centred around [RA; DEC] = [214.477° ;  -45.411°] }
\label{fig:2mass}
\end{figure}

Figure \ref{fig:2mass}a presents the comparison between the magnitudes extracted in our image for the 97.5\% of good detections with those of the \textit{2MASS} catalog. 
This figure exhibits few characteristics features: (i) an overall good agreement with \textit{2MASS} magnitudes and (ii) an underestimation of the flux of some faint stars with magnitude J $\geq 15$, and (iii) a slight over-estimation of the flux for the brightest stars. We discuss these three points below:

\begin{itemize}
    \item[(i)] The magnitudes measured on the preprocessed image globally agree with the input magnitudes. The difference has a median $\Delta \mathrm{J} = 0.11$ mag and a standard deviation of 0.33 mag. This difference is due to the choice of a crude zero-point that is not precise at 0.1 magnitude level. 
    \item [(ii)]After verification, these stars are located on top of a cold or dead pixel. The cold pixel reduces the signal measured for the star. This effect will be properly accounted for by the astronomy pipeline, using the calibration map of cold pixels.
    \item [(iii)] The brightest stars include pixels reaching saturation within 1 or 2 frames. This makes the evaluation of the slope difficult for these pixels, and the signal estimation is less precise. 
\end{itemize}

\begin{figure}[h]
    \centering
    \includegraphics[width=10cm]{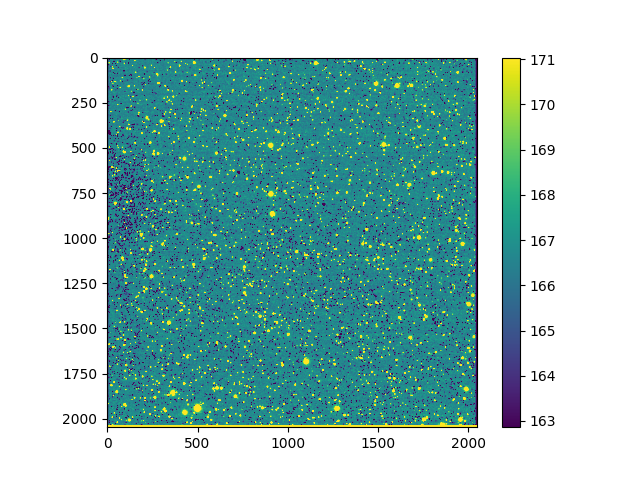}
    \caption{Signal map of a sky region centred around [RA; DEC] = [214.477° ;  -45.411°], with a higher star density}
    \label{fig:champs2}
\end{figure}

Figure \ref{fig:2mass}b presents the comparison between the magnitudes extracted for a sky region with a higher star density. 
The field of view is centred around [RA; DEC] = [214.477° ;  -45.411°]. A signal map of this sky region is given figure \ref{fig:champs2}.  Without persistence effect from previous acquisition, we obtain here 100\% of good detections.
The blue points show the magnitudes obtained with no correction of the cold pixels (point (ii) discussed above). The orange points show the impact of a simple correction that replaces the signal of Cold pixels by the median signal of the 8 pixels surrounding them. Interestingly, this simple correction significantly improves the situation, giving useful indication for the astronomy pipeline. 
The conclusions are the same as for Figure \ref{fig:2mass}a, except for the stars located on top of cold pixels whose signal has been corrected. Moreover, the three stars presenting an underestimated magnitude are located at the edge of the detector, so that their signal is truncated. 
This second field of view, (fig. \ref{fig:champs2}) will be our reference for the following sections. 

In the following sections, we discuss the impact of some key steps of the Preproc, for the construction of the preprocessed image.

\subsection{Measuring the signal before the saturation}
\label{sub:sat}
 
\begin{figure}[h]
    \centering
    \includegraphics[width=\columnwidth]{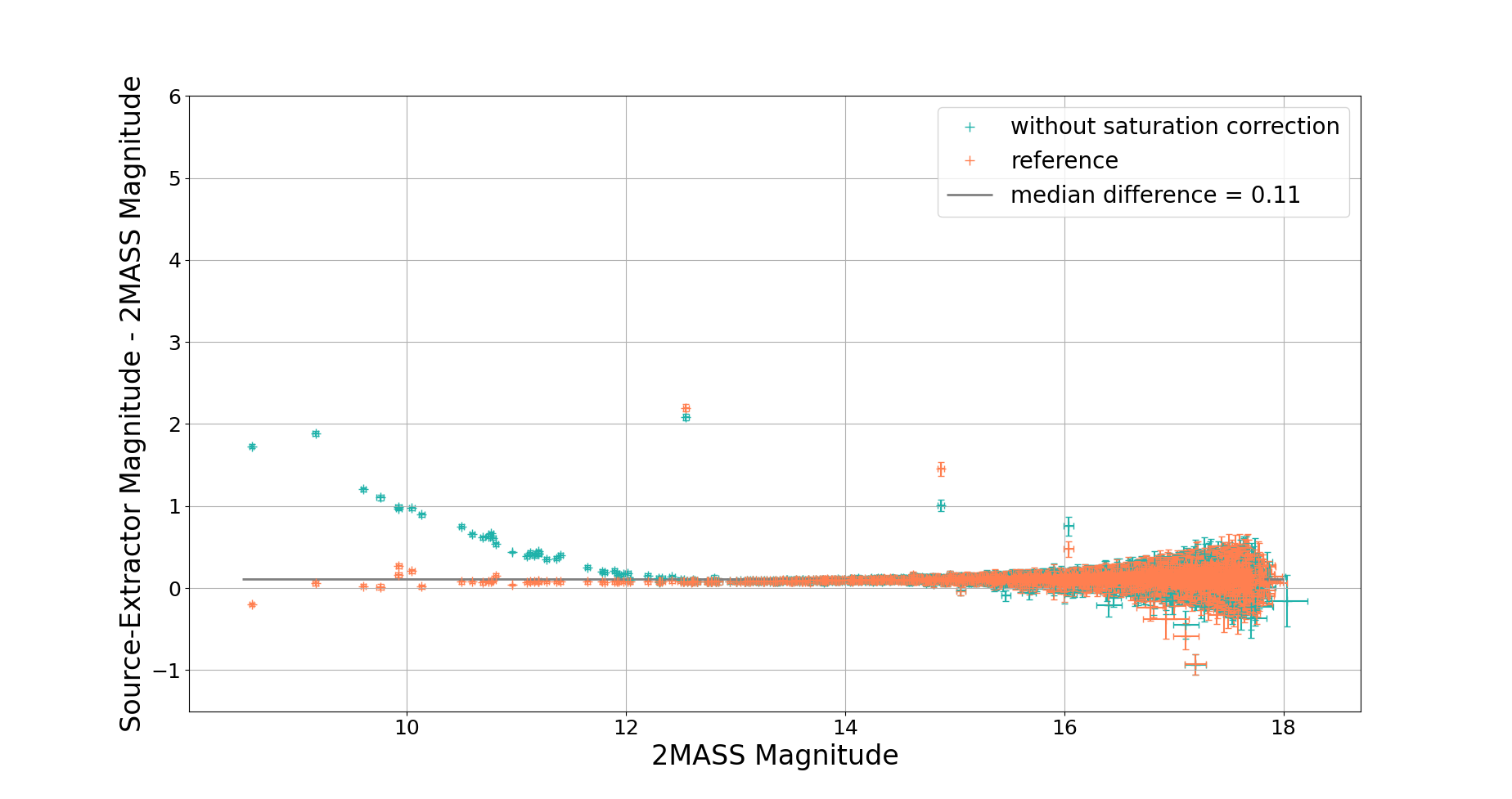}
    \caption{Difference between magnitudes estimated by Source-Extractor and \textit{2MASS} magnitudes as a function of \textit{2MASS} magnitudes. Orange points are the results  computed with the complete pipeline, presented figure \ref{fig:2massB}. Blue stars are the magnitudes computed with saturated pixels set to saturation. The blue point below magnitude 12 reflect an underestimation of the flux linked to not considering the saturation level in the computation }
    \label{fig:dif_sanssat}
\end{figure}

Measuring the signal every 1.33\,s allows to compute the signal of bright sources with only the unsaturated part of the ramp. We have chosen to fit the differential ramp up to 70\% of the saturation level, for each pixel. 
To demonstrate the relevance of this approach, we compare in figure \ref{fig:dif_sanssat} the magnitudes estimated with and without knowing the frame at which the signal saturates. 
This figure shows that the knowledge of the frame at which the signal saturates allows to recover the magnitude of bright stars in the range J $\approx 9 - 12$, whose flux is significantly underestimated otherwise. 

\subsection{Correcting non-linearities}
\label{sub:pnonlin}
\begin{figure}[h]
    \centering
    \includegraphics[width=\columnwidth]{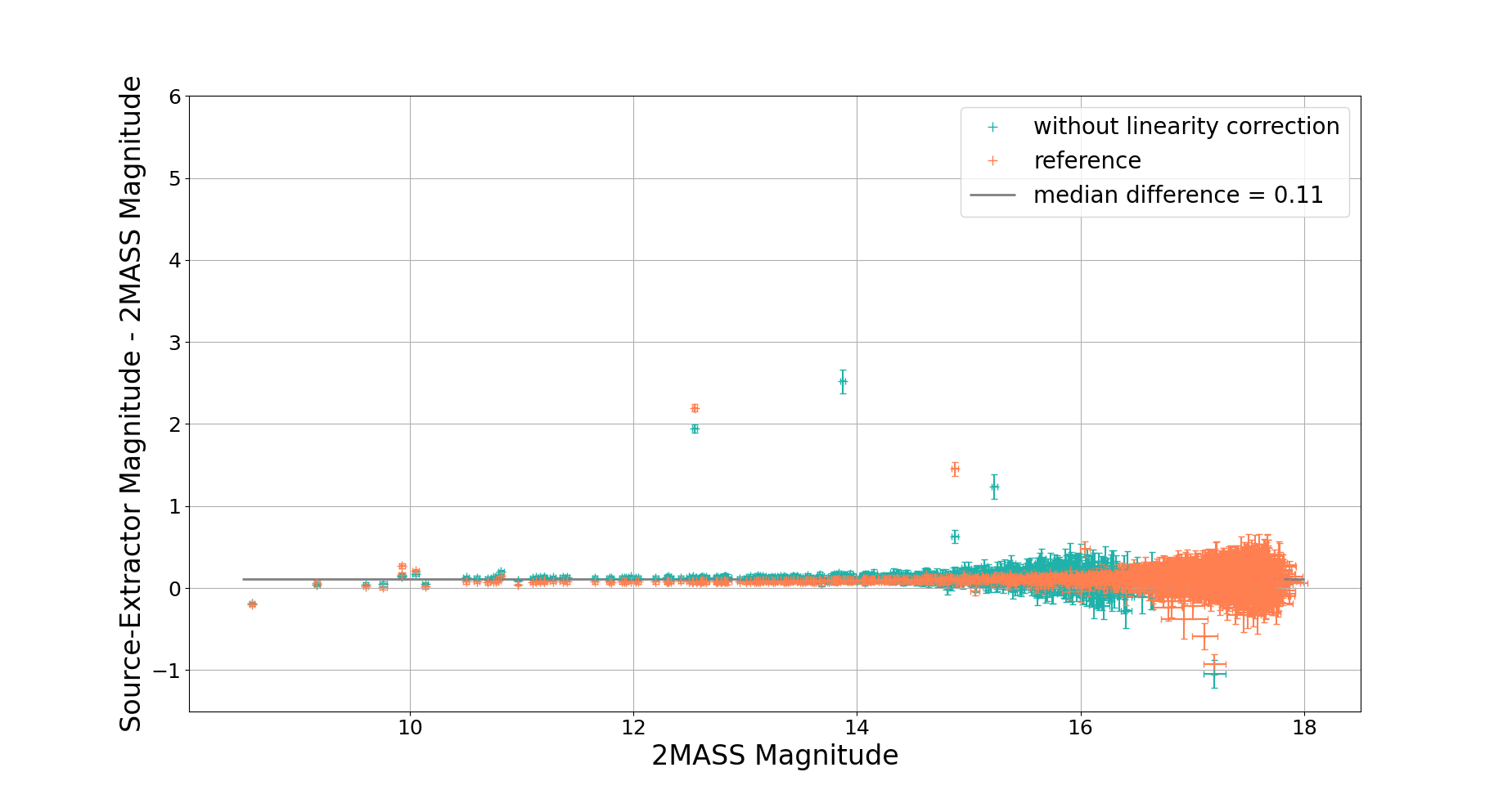}
    \caption{Difference between magnitudes estimated by Source-Extractor and \textit{2MASS} magnitudes versus magnitudes referenced in \textit{2MASS}. Orange points are the results computed with the complete pipeline, presented in figure \ref{fig:2massB}. Blue stars are the magnitudes computed on the signal map computed as the median value of the ramp}
    \label{fig:dif_sanslincor}
\end{figure}

\begin{figure}[h]
    \centering
    \includegraphics[width=\columnwidth]{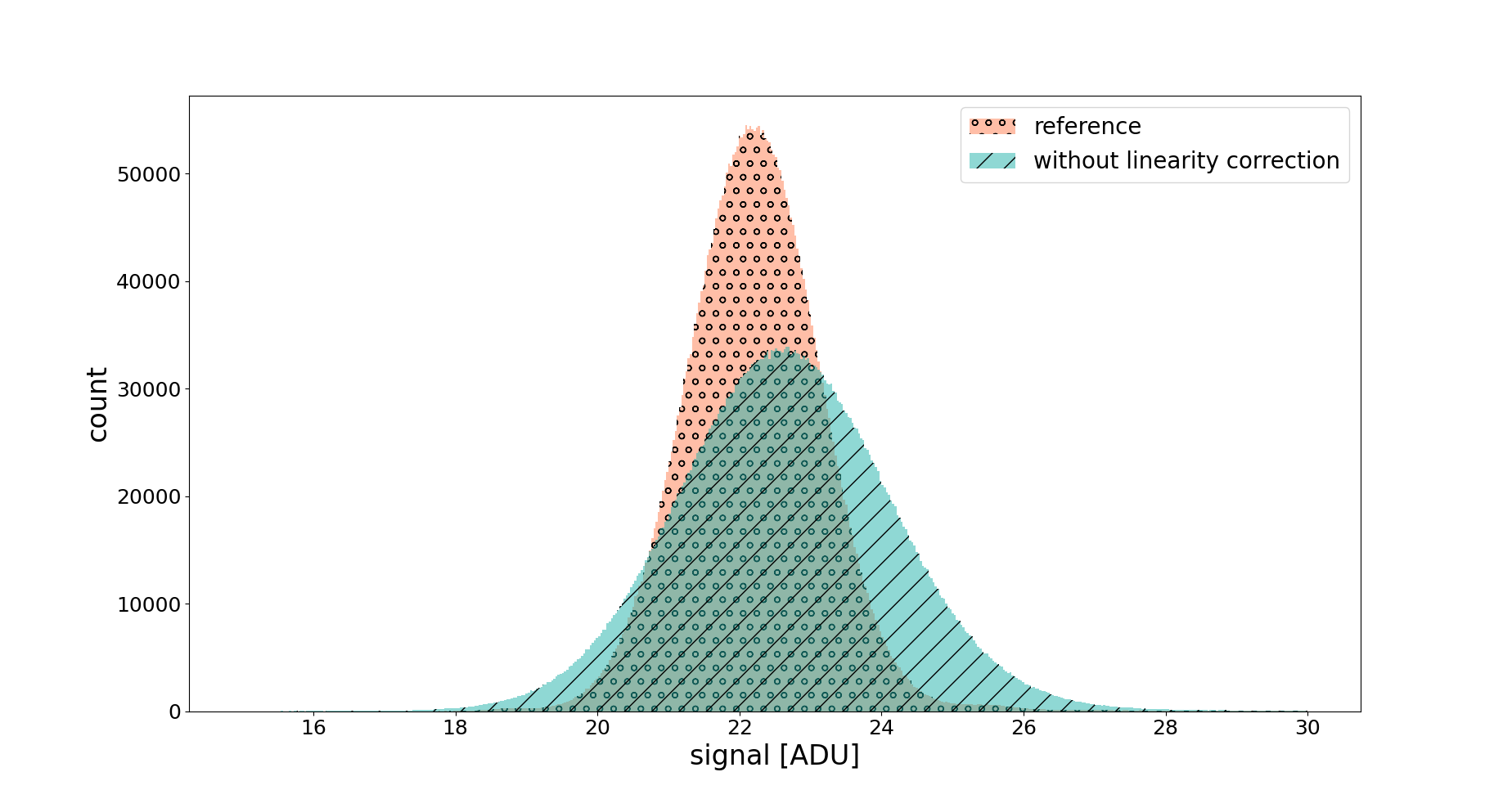}
    \caption{Histogram of the sky signal (slope of the ramp), measured with and without non-linearity correction. 
    The orange histogram (with circles), refers to the signal measured by the pre-processing pipeline, including the non-linearity correction described in \ref{sub:nonlin}. For the blue hatched one, the signal is simply computed as the median of the differential ramp.  }
    \label{fig:hist_nonlin}
\end{figure}

The ramp non-linearities are corrected thanks to the non-linearity coefficient defined in section \ref{sub:maps}. 
Figure \ref{fig:dif_sanslincor} compares the magnitude reconstruction with (orange points) and without (blue points) considering the ramp non-linearities. 
The points show the difference between magnitudes computed from the simulated CAGIRE ramp and \textit{2MASS} magnitudes, as a function of \textit{2MASS} magnitudes. The sky signal measured on the uncorrected map is more dispersed (see figure \ref{fig:hist_nonlin}), leading to a loss of sensitivity to faint stars. Indeed, with our crude parametrization of Source Extractor, 60\% of the stars detected beyond magnitude J = 16, are not detected when we do not make the non-linearity correction.

%The orange (circles filled) histogram is the signal histogram with correction of the non-linearities, included in the preprocessing pipeline. It is the signal used to compute the magnitudes plotted in orange in figure \ref{fig:dif_sanslincor}. The blue (hatched) histogram is the signal histogram without correction of the non-linearities. The signal is here computed as the median of the differential ramp. It is the signal used to compute the magnitudes plotted in blue in figure \ref{fig:dif_sanslincor}

\subsection{Impact of the flatfield correction}
\label{sub:flatfield}

Before computing the magnitudes, the signal map delivered by the preprocessing pipeline must be corrected for the flatfield. 
This correction is obtained by dividing the signal map by a previously measured flatfield map.
The detector flatfield map is computed with a ramp acquired under uniform and constant illumination. The signal measured in each pixel is divided by the median signal over the entire detector, to make the flatfield map. 

\begin{figure}[h]
    \centering
    \includegraphics[width=\columnwidth]{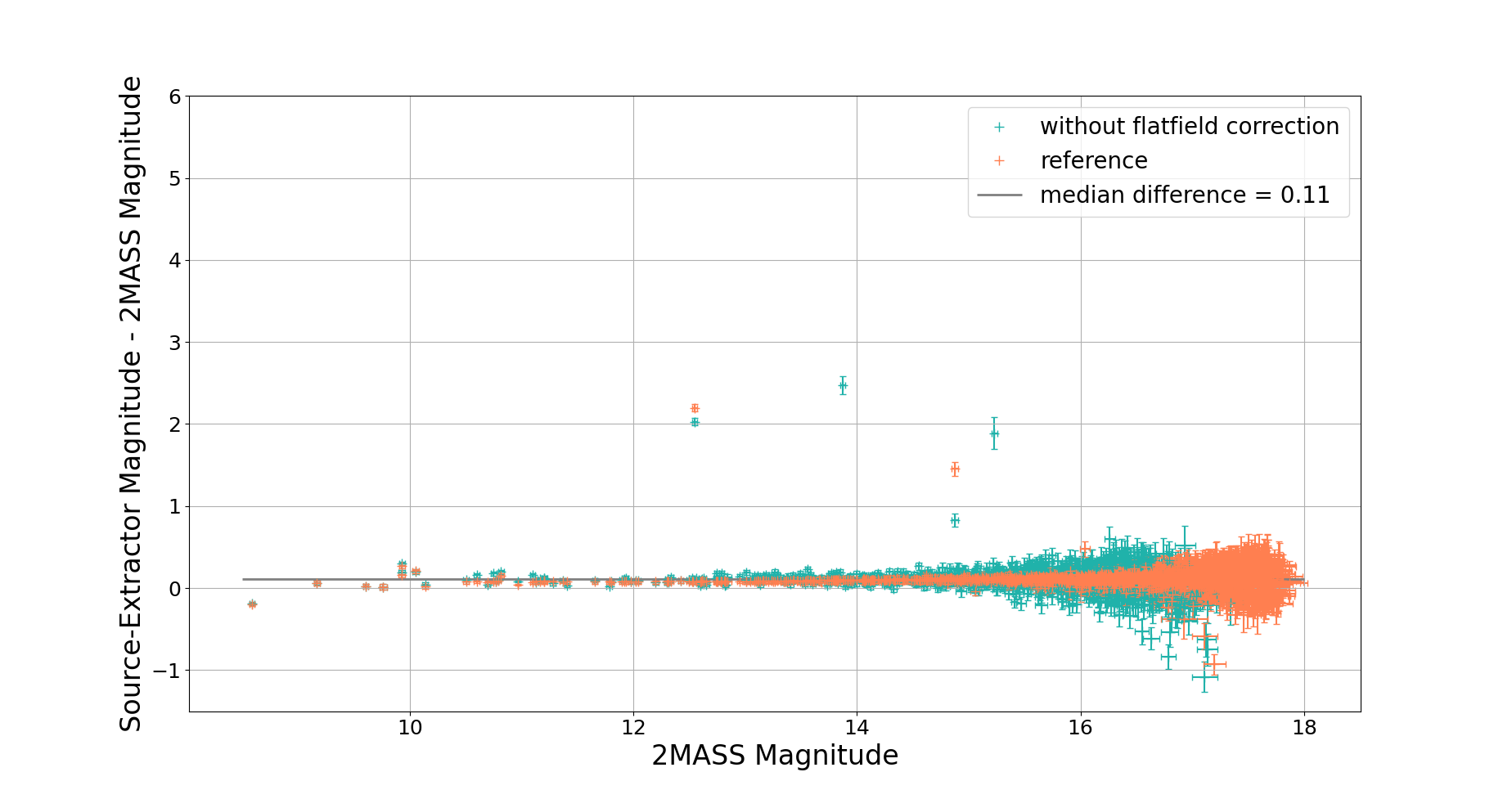}
    \caption{Difference between magnitudes estimated by Source-Extractor and \textit{2MASS} magnitudes as a function of \textit{2MASS} magnitudes. Orange points are the results computed with the complete pipeline, presented figure \ref{fig:2mass}b. Blue stars are the magnitudes computed on the signal map which is not corrected by the flatfield}
    \label{fig:dif_sansflatcor}
\end{figure}

Figure \ref{fig:dif_sansflatcor} compares the magnitudes computed with (orange points) and without (blue points) the division by the \textit{detector} flatfield. 
The points show the difference between magnitudes computed from the simulated CAGIRE ramp and \textit{2MASS} magnitudes, as a function of \textit{2MASS} magnitudes. 
Figure \ref{fig:dif_sansflatcor} shows that for magnitudes below J = 17, the magnitudes seem well estimated, but beyond this magnitude small stars may be missed if the detector flatfield is not taken into account. 
116 stars are recovered in the signal map without flatfield correction versus 192 when the flatfield correction is applied. 
Moreover, the absence of flatfield correction leads to 4 "false detections", which disappear when the flatfield correction is taken into account. 
Flatfield correction is thus an essential step before star extraction.

Of course, the flatfield correction performed here is not representative of the real situation, with CAGIRE at the focus of COLIBRI, because the flatfield will be very different and it will be processed by the astronomy pipeline, which comes after the Preproc. 
We have discussed it here to illustrate the impact of the sole detector non-uniformities.

Finally, we note that the inter-pixel capacitance is not taken into account by the Preproc. Indeed, its impact on magnitude estimation has been studied with the conclusion that the cross-talk has no effect on the source detection and the measure of the magnitude. 
The mean difference between the magnitudes computed with cross-talk and without cross talk is only -0.02 mag, with a standard deviation of 0.1 mag. Compared to other effects, the cross talk has thus a marginal impact on the image quality.

The analysis of simulated ramps shows the ability of the Preproc to measure properly the signal received by each pixel, after correction of several effects, such as the detector non-linearity, the saturated pixels and pixels impacted by a cosmic ray. We expect the Preproc to provide reliable signal maps for the astronomy pipeline. 

In the next section, we address the impact of signal persistence, and the possibility to take it into account in the Preproc.

\subsection{Impact of persistence on source detection}
\label{sub:ppersistence}
  
 While persistence in MCT detectors is a truly complex phenomenon, involving various physical processes \citep{Legoff2020}, the characterisations made by CEA-IRFU have shown that it can be satisfactorily modeled with the sum of 3 exponentials (see \cite{Legoff2020} and section \ref{sub:persistence}). 
 According to this model, the persistence from bright stars appear as faint decaying signals which may be confused with a GRB afterglow. 
 Within the context of preprocessing CAGIRE images, we use this model to study the impact of saturating stars on subsequent images. 
 Considering the rapid decay of the persistent signal illustrated in figure \ref{fig:persistance_signal}, we specifically study the impact of one ramp on the following one.
 This limited study aims at verifying whether the persistent signal from a bright star can be detected in the short duration of a ramp, and eventually if a mitigation procedure can be constructed for persistence.
  
 \subsubsection{The simulation}
\label{ssub:simuper}

We simulate two ramps with a varying time delay between the end of the first ramp and the beginning of the second one. 
The two ramps cover the same field of view, with a shift by +0.04~/~-0.12 degrees in RA / DEC for the second ramp (+222~/~-665 pixels).
%The two ramps cover the same field of view, with a shift by +0.16~/~-0.05 degrees in RA / DEC for the second ramp (+886~/~-277 pixels).
 The persistence from the first ramp is modelled as explained in section \ref{sub:persistence}.
The usual Preproc is then run on the second ramp, with or without persistence mitigation (cf. section \ref{sec:preproc}, and we check whether the persistent signal from the first ramp leads to false detections in the second ramp. To mitigate the persistence, we follow the procedure explained in section \ref{sec:preproc} to subtract the persistent signal to pixels which saturated in the previous acquisition. 

 We simulate various cases: 
\begin{itemize}
    \item  Without persistence.
    \item  With persistence and zero delay between the 2 ramps.
    \item  With persistence and a variable delay between the 2 ramps, ranging from 1\,min to 10\,min.
    \item With persistence, zero delay between the 2 ramps, and a mitigation procedure.
    \item With persistence, a variable delay between the 2 ramps, and a mitigation procedure.
\end{itemize}

\begin{figure}[h]
    \centering
    \includegraphics[width=\columnwidth]{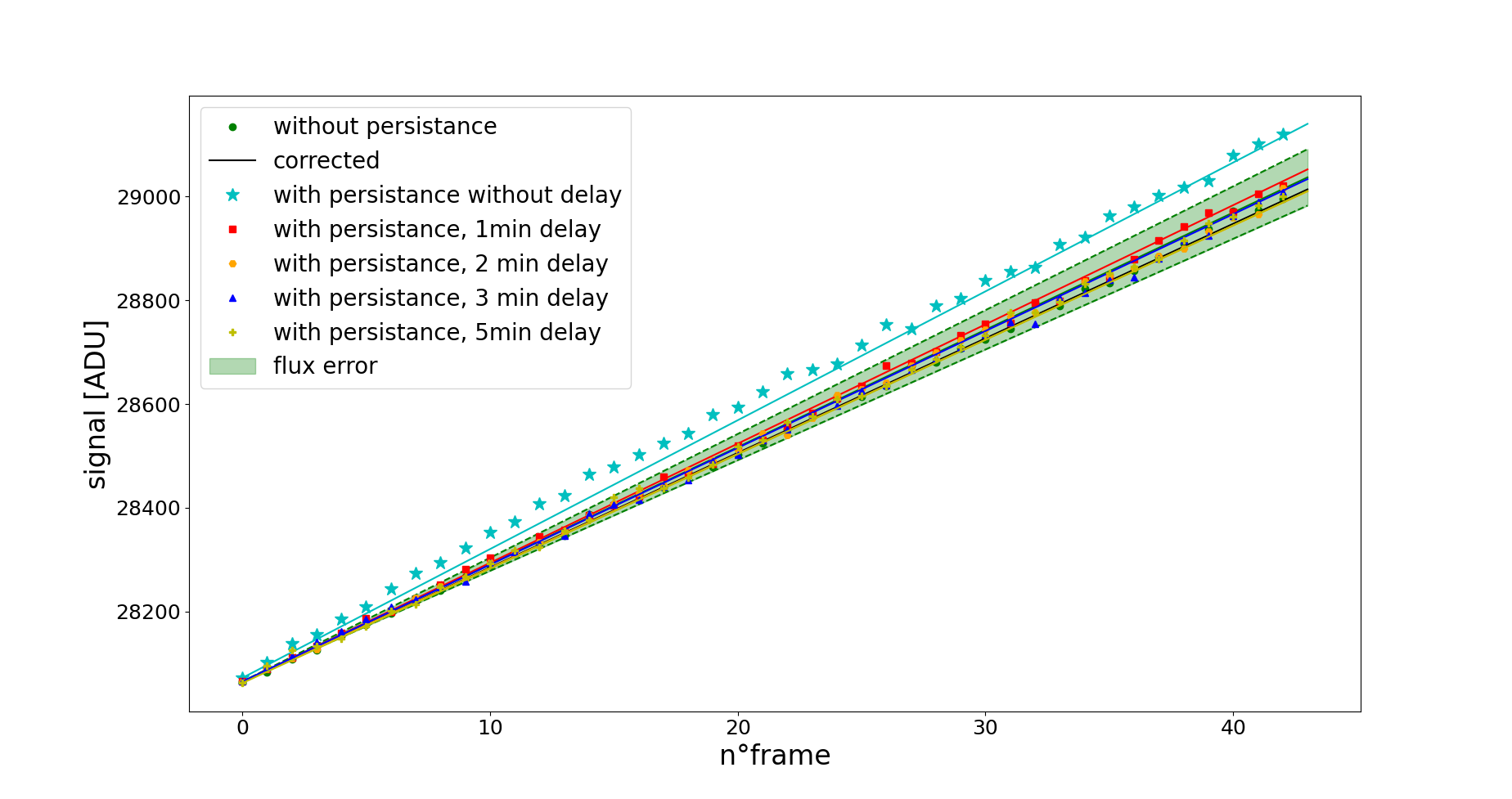}
    \caption{Simulated ramps (symbols) and estimated signal (full lines) for a pixel impacted or not by persistence. Cyan stars represents the pixel signal impacted by persistence with no delay between the 2 ramps. Red squares, orange points, blue triangles and yellow crosses represent the signal of the same pixel impacted by persistence with a variable delay between the two ramps. Finally, green circles represent the signal without persistence and the green region the 1$\sigma$ error on its estimation}
    \label{fig:rampesP}
\end{figure}

Figure \ref{fig:rampesP} presents the simulated ramps for these 4 cases for a pixel illuminated by a bright star in the first ramp. We see that, after 1\,min, the signal added by persistence has reduced considerably and is within the error bars. 
This trend has been confirmed on 4 others pixels, presented in table \ref{tab:pourcentagePersistance}, which shows the relative impact of persistence on the pixel's signal compared with the error on the signal estimation. 

This study points out that after 1 min, the impact of the persistent signal is smaller than the 1$\sigma$ error of the signal estimation for on a single ramp. With a maximum illumination level of 200\%, persistence from a ramp should impact strongly only the ramp immediately following. 
However, it also shows that for two consecutive ramps, uncorrected persistence can lead to false detections. 
The mitigation procedure, on the other hand, leads to correct signal estimates, as shown in table \ref{tab:pourcentagePersistance}. 
Cross matching the sources identified in our simulated images with \textit{2MASS} does not show unmatched detections when we apply the persistence correction. 
To confirm this observation, the relative difference between the corrected signal and the signal without persistence is indicated in the last column of table \ref{tab:pourcentagePersistance}.

\begin{table}[h]
    \centering
    \begin{tabular}{|c|c c c c c||c c c|}
    \hline
    delay &  0min & 1min& 2min& 5min &10min  &  0min &  1min &  2min \\
      &   \multicolumn{5}{c||}{before correction} &  \multicolumn{3}{c|}{after correction }  \\
    \hline
    pixel n°1  & \textbf{21.4\% } &  \textbf{ 7.95\%} & 4.0\%   &  5.04\%  &  1.13\%   &- 1.3\% & 1.6\% & 0.2\% \\
    \hline
    pixel n°2 &\textbf{  8.2\% } &  2.67\% & 3.17\%  &  1.55\%   & -0.47\%    & -1.7\% & -0.2\%   & 2.5\% \\
    \hline
    pixel n°3 & \textbf{ 9.1\% } &  1.55\%  & 1.25\%  &   1.29\%  & -0.4\%   &- 1.4\%& -0.3\%  & 0.1\% \\
    \hline
    pixel n°4 & \textbf{ 9.9\%}&1.69\%  & -2.67\%   & -2.4\%   & -3.41\%  &-2.3\% & -0.1\% & -3.7\%\\
    \hline
    pixel n°5  &\textbf{13.2\%  } & 4.43\% &   3.60\%  &  1.33\%  & -0.79\%  &  -1.7\%& 2.6\% & 2.5\%  \\
    \hline    
    \end{tabular}
    \caption{Relative difference between the signal measured with and without persistence, for different pixels and different time delays between two ramps. Values in bold exceed the 1 sigma error on the signal estimation. After correction, the relative difference is well below the $\sim$6\% relative error on the signal computation (1 sigma)}
    \label{tab:pourcentagePersistance}
\end{table}

Based on these simulations, we are thus optimistic on the capability of the Preproc to provide reliable signal maps corrected from the persistent signal of the previous acquisition.  

\subsection{Summary}
\label{sub:conclusion_simu}
The construction of realistic simulated ramps was an important step to validate the pre-processing pipeline. 
We have shown that the preprocessing can efficiently take into account detector effects, such as the pixel saturation, the ramp non-linearities and flatfield corrections to produce reliable maps of the signal received by the detector. These simulations also stressed the need to take into account the persistence from previous exposure(s) to avoid detecting false stars.

While the simulations permit the end-to-end tests of the Preproc and an evaluation of its performance, they represent an ideal case with respect to the real detector looking at a real sky through a real telescope. 
To deal with this situation, the Preproc contains a number of adjustable parameters that will be tuned during the commissionning phase of the instrument, in order to provide clean signal maps to the astronomy pipeline.

\section{Conclusion}
\label{sec:conclusion}

We have presented the CAGIRE camera, a NIR wide-field imager dedicated to the follow-up of high-energy and multi-messenger transients in the near infrared, at the focus of the COLIBRI 1.3 meter robotic telescope. In the context of the \svo\ mission, CAGIRE will provide a crucial link between the localizations of ECLAIRs  and large ground-based telescopes. 

The location of CAGIRE at the focus of a robotic alt-azimutal telescope involves specific adaptations, which justify the camera design. We have presented the camera optical, mechanical and electric architecture, highlighting the characteristics of the ALFA detector and its acquisition chain. The first characterizations presented here have allowed to develop an image simulator for CAGIRE, that we have used to test the preprocessing pipeline. These tests have illustrated the ability of the Preproc to correct effects such as non-linearity, flatfield, saturation or persistence. A particular emphasis has been put on persistence correction, as it could lead to "false detections". We also stressed the importance of the flatfield correction for the identification of faint sources. 

CAGIRE is expected to be on the sky mid-2024, after additional characterization of the detector at CPPM, the integration of the camera at IRAP followed by a series of tests, and the final shipment of the instrument to Mexico.

\subsection*{Acknowledgments}

CAGIRE is partly funded by the French Centre National d’Etudes
Spatiales (CNES). 
The PhD contract of A. Nouvel de la Flèche is financed by CNES and LYNRED. 
The development of the ALFA detector has been funded by ESA contract 22949/09/NL/CP. We are grateful to ESA for the loan of an ALFA detector to the CAGIRE project\footnote{The view expressed herein can in no way be taken to reflect the official opinion of the European Space Agency.}.
This work has been partially supported by the LabEx FOCUS ANR-11-LABX-0013.
The calibration data used in the ALFA detector section of this paper were acquired at CEA.
We acknowledge the Observatorio Astronómico Nacional and the Instituto de Astronomía of the Universidad Nacional Autónoma de México for providing data acquired with the RATIR instrument for the validation of the Preproc. We particularly thank Alan Watson for the data acquisition and Nathaniel Butler for useful advice on the preprocessing.
We would like to thank two IRAP engineers who had a significant contribution to CAGIRE before their retirement: Francis Beigbeder and Patrick Couderc.

Softare: Matplotlib \citep{Hunter2007}, numpy \citep{Harris2020}.

\section{Declarations}
\subsection{Financial statement}
The PhD contract of A. Nouvel de la Flèche is financed by CNES and LYNRED. 
The development of the ALFA detector has been funded by ESA contract 22949/09/NL/CP.
This work has been partially supported by the LabEx FOCUS ANR-11-LABX-0013.

\subsection{Conflict of interest}
The authors declare that they have no conflict of interest.

\bibliography{sn-bibliography.bib} % common bib file
%\input sn-article.bbl

%\input sn-sample-bib.tex%

\begin{appendices}

\section{ appendix}
\label{sec:annexe}

\subsection{Diagram of the overall electronics architecture}
\label{sub:A1}

\begin{figure}
    \centering
    \includegraphics[width=12cm]{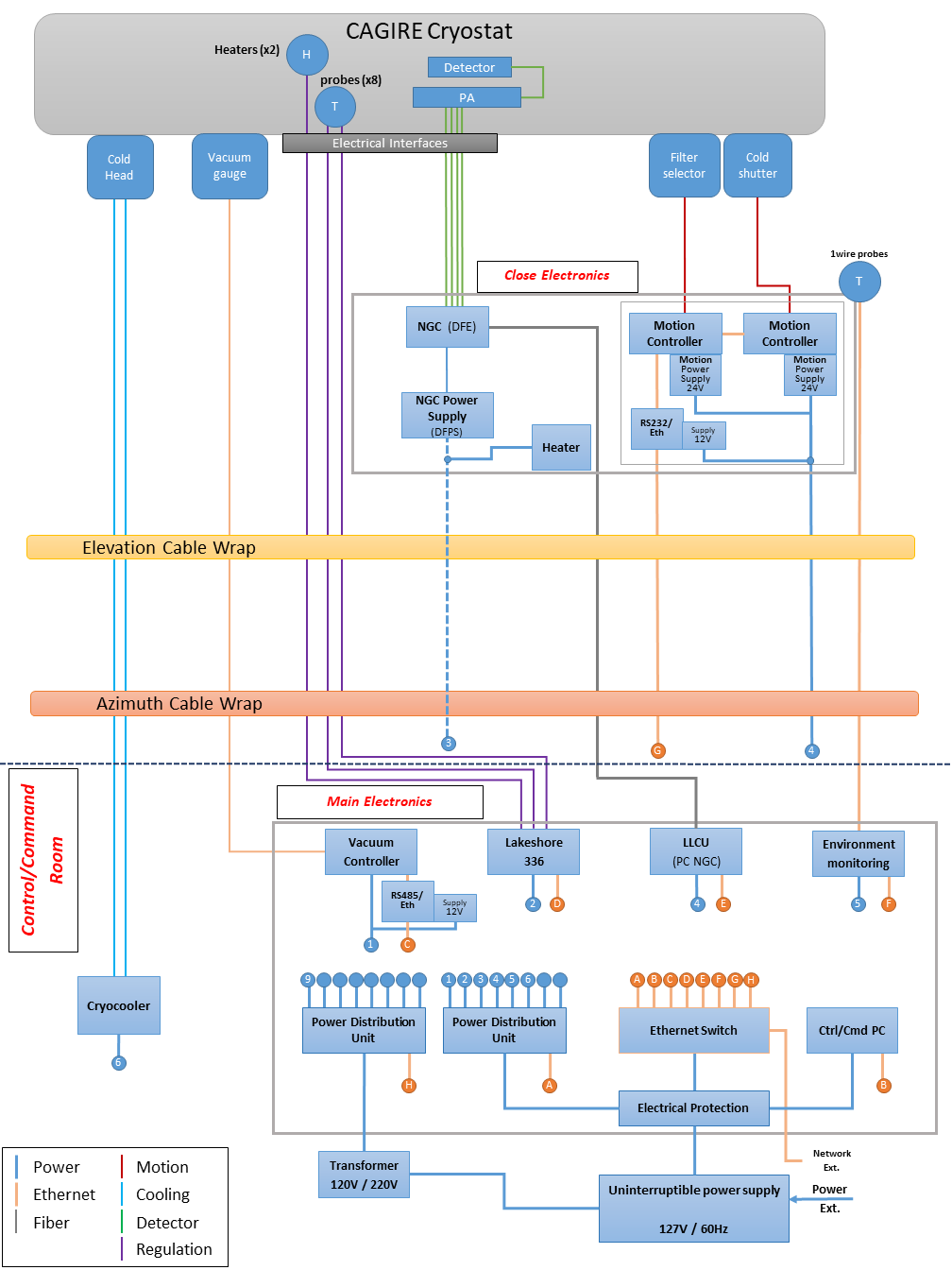}
    \caption{Diagram of the overall electronics architecture}
    \label{fig:electronics}
\end{figure}




\end{appendices}


\end{document}